\documentclass[a4paper,twocolumn,11pt,accepted=2021-07-07]{quantumarticle}
\pdfoutput=1

\usepackage[numbers,sort&compress]{natbib}
\usepackage{color}
\usepackage{bbm}
\usepackage{graphicx}% Include figure files
\usepackage{dcolumn}% Align table columns on decimal point
\usepackage[utf8]{inputenc}
\usepackage{graphicx}
\usepackage{epstopdf}
\usepackage{amsthm}
\usepackage{cancel}
\usepackage{amsmath}
\usepackage{empheq}
\usepackage{bbm}
\usepackage{braket}
\usepackage{amssymb}
\usepackage[usenames,dvipsnames]{xcolor}
\usepackage{cases}
\usepackage{latexsym}
\usepackage[colorlinks=true,citecolor=Cerulean,linkcolor=RubineRed,urlcolor=Cerulean]{hyperref}

\newcommand{\be}{\begin{equation}}
\newcommand{\ee}{\end{equation}}
\newcommand{\ba}{\begin{eqnarray}}
\newcommand{\ea}{\end{eqnarray}}
\newcommand\tr{{\mbox{Tr\,}}}
\newcommand{\ignore}[1]{}

\begin{document}

\title{Energy storage and coherence
in closed and open quantum batteries}

\author{Francesco Caravelli}
\affiliation{Theoretical Division (T4),\\
Los Alamos National Laboratory, Los Alamos, New Mexico 87545, USA}
\orcid{0000-0001-7964-3030}
\thanks{Corresponding author}
\email{caravelli@lanl.gov}
\author{Bin Yan}
\affiliation{Center for Nonlinear Studies,\\
Los Alamos National Laboratory, Los Alamos, New Mexico 87545, USA}
\affiliation{Theoretical Division (T4),\\
Los Alamos National Laboratory, Los Alamos, New Mexico 87545, USA}
\orcid{0000-0003-2258-2817}
\author{Luis Pedro Garc\'ia-Pintos}
\affiliation{Joint Center for Quantum Information and Computer Science and Joint Quantum Institute, NIST/University of Maryland, College Park, Maryland 20742, USA}
\orcid{0000-0003-2075-4996}

\author{Alioscia Hamma}
%\email{alioscia.hamma@umb.edu}
\affiliation{Department of Physics, University of Massachusetts, Boston, Massachusetts 02125, USA}
\orcid{0000-0003-0662-719X}

\maketitle

\begin{abstract}
We study the role of coherence in closed and open quantum batteries. We obtain upper bounds to the work performed or energy exchanged by both closed and open quantum batteries in terms of coherence. Specifically, we show that the energy storage can be  bounded by the Hilbert-Schmidt coherence of the density matrix in the spectral basis of the unitary operator that encodes the evolution of the battery. We also show that an analogous bound can be obtained in terms of the battery's Hamiltonian coherence in the basis of the unitary operator by evaluating their commutator.  We apply these bounds to a  4-state quantum system and the anisotropic XY Ising model in the closed system case, and the Spin-Boson model in the open case.
\end{abstract}

\section{Introduction}
With the improvement of  technology able to manufacture and manipulate solid-state devices, we are now able to harness the properties of physical systems at the nanometer scale~\cite{alicki0,Kosloff,levy,alicki,LindenPRL2010,campaioli,Bhattacharjee}. In this regime, their behavior can be affected by quantum phenomena, and thermodynamic laws have to be investigated in detail~\cite{AlhambraPRX2016,2019arXiv190202357A,masanes2017general,workextractionBrandaoPRL13,gallego2016thermodynamic,Huberreviewthermo2016,Kosloff2014,horod,sss,AdC2014,Funo17,lostaglionature,Nieuwenhuizen,janzig,horodecki, brandao2,lostaglio2, cwiklinski,lostaglio3,bera,wilming,muller,gour,streltsov,kwonwork}. This includes physical systems designed to store energy~\cite{aaberg2013truly,polini,PoliniPRB2018,PoliniPRL2019,PoliniPRB2019,andolinasyk,Caravelli1,LewensteinBatteries18,Modiarxiv2017,GelbwaserNJP2015,FrenzelPRE2014,HuberGaussianBatteries2017,adc,adc2,battp}. Genuine quantum phenomena are connected to either some interference pattern, or to the incompatibility of different observables. Such notions are unified within the framework of {\em quantum coherence} \cite{workextractionAbergPRL14,brandner2015coherence,FrancicaPRE2019,korzekwa2016extraction,PetruccioneSciRep2019,lostaglio2,zanardicoh,brac,MarvianPRA2016,plenio,streltsov}, that, in simple words, is a way of evaluating the anticommutability of a given observable with the state of a system~\cite{plenio}. 
Quantum coherence can be also described in term of resource theory \cite{anders2017focus,workextractionPopescuNatComm14,oppenheim,winter}.
%,as it is of fundamental importance for a number of quantum technologies such as quantum metrology, quantum communication, and quantum 
%thermodynamics. 
Since resource theories serve a the bedrock of thermodynamics~\cite{workextractionBrandaoPRL13}, it is natural that quantum coherence and the role of entanglement~\cite{sss,erg,lis,hammaavg,hammaavg2,horod,quach} have also been thoroughly studied in the context of quantum thermodynamics~\cite{lostaglio2}. 

%%%%%%%%%%%%%%%%%%%%%%%%%%%%%%%%%%%%%%%%%%%%%%%%%%%%%%%%%%%%%%%%%%%%%

Among the quantum devices capable of performing work, quantum batteries have a special place. Quantum batteries are of fundamental importance, and are an area of intense study~\cite{alicki0, aaberg2013truly, GelbwaserNJP2015, AlhambraPRX2016, 2019arXiv190202357A, levy, AdCPRL2017cycleengine, masanes2017general, CorreaPRE2013,anders2017focus, LindenPRL2010, workextractionPopescuNatComm14}, in thermodynamics~\cite{campaioli, alicki, polini, PoliniPRB2019, PoliniPRB2018, PoliniPRL2019, Modiarxiv2017, LewensteinBatteries18, Caravelli1}. 
We model a quantum battery by a Hamiltonian $H_0$ that gives a notion of energy, and a quantum state $\rho_t$ evolving in time as $\mathcal E_t(\rho) = \rho_t$. Here the map $\mathcal E_t$ is a generic quantum channel that needs not to be unitary, as we consider also the possibility of open quantum systems~\cite{workextractionPopescuNatComm14,luis,crescente}.
%by means of a quantum evolution.
The extracted or stored work results from populating the levels of $H_0$ in a different way from the initial state. %Notice that the quantum evolution superoperator $C_t$ needs not to be unitary, as we consider also the possibility of open quantum systems~\cite{workextractionPopescuNatComm14,luis}.
%It is not clear in general  how efficiently nano-batteries can operate \cite{anders2017focus, LindenPRL2010, workextractionPopescuNatComm14}, and what are the limits to speed of energy extraction (the power) or the total amount energy that can be stored or extracted \cite{luis}. 

Previous work has shown the importance of quantum coherence in extracting work from a quantum system. In~\cite{kwonwork}, the authors studied the amount of coherence in the eigenbasis of the Hamiltonian that can be extracted by a thermal process. Meanwhile,~\cite{lostaglionature} shows how the behavior of quantum coherence poses further constraints to thermodynamic processes, in particular, coherence transformations are always irreversible.  
There are already other various bounds on the performance of quantum batteries~\cite{LewensteinBatteries18,CorreaPRE2013,AdCPRL2017cycleengine,levy,luis} and recently some appeared in which, specifically, coherence has been investigated \cite{Plastina,cakmak}, and in particular $l_1$ coherence \cite{cakmak2}.

In this paper, we obtain quantitative {and rather general} bounds on work (extracted energy), and power in both closed and open quantum systems in terms of the  coherence of both the state or the Hamiltonian $H_0$ in the  eigenbasis of the time evolution operator. By coherence of Hamiltonian in a given basis we mean the amount of non commutativity of the Hamiltonian in that basis and discuss in what regimes the bounds are tight. We show applications to several systems of physical interests: arrays of closed quantum batteries, a  spin chain after a quantum quench, and an open system described in the spin-boson setting.

The paper is organized as follows. In Section II we introduce the coherence operators which we use in the paper, and obtain upper bounds to the work, power in terms of the coherence of the battery Hamiltonian and the density matrix in the basis of the evolution operator.
%We show that these bounds are exactly zero for decoherence free substates. 
In Section III we show how these results can be generalized to generic quantum channels, by finding bounds in terms of both Kraus and Lindblad operators connecting the energy exchange to the coherence in the basis of these operators.
Finally, in Section IV we show applications of these bounds. For the closed quantum system case, we consider
a 4-level system composed of two interacting spins, and a many body system given by the anisotropic XY model. We show that our bounds are fairly %\LP{fairly seems a bit vague: do we mean something like within a certain factor, or something like that?}
tight for small systems in which coherence is important.
%In the XY model, we find that the bounds are tight for small quantum quenches. 
For the open system case, we consider a spin-boson model, showing that the coherence bounds give the right dependence at short times.
Conclusions follow.
%\LP{here we are referring to the total energy change, right? Are bounds on power close to tight then?}.  

\section{Coherence and work in closed quantum batteries}
\subsection{Energy storage and coherence bounds}
We describe a closed quantum battery by a quantum system on a Hilbert space $\mathcal H\simeq \mathbb{C}^n$ with time dependent Hamiltonian $H(t)=H_0+V(t)$. Time evolution generated by the Hamiltonian is unitary and reads $\dot{\rho}_t=i[H,\rho_t]$, where we use units in which $\hbar=1$. 
The energy of the battery is measured as the expectation value of the bare Hamiltonian $H_0$ so that the {work} extracted from the battery is  \cite{campaioli}
\begin{equation}
{W(t)=\text{Tr}\Big((\rho_0-\rho_t)H_0\Big)},
\end{equation}
where we used {the fact that} no entropy change occurs for unitary dynamics. {
This kind of quantum battery is described by a closed time-dependent quantum system. 
%these systems 
The battery is externally driven by $V(t)$ in order to modify its population levels and change its energy.} 
We derive upper bounds to the energy, and thus also find maximum bounds on the \textit{ergotropy}, which is the maximum energy change when maximizing over unitary operations \cite{campaioli}, and in particular of the \textit{isospectral twirling} of the work, e.g. the spectral-preserving unitary average over the time evolution \cite{Caravelli1,isospec1,isospec2}, or entangling power \cite{adpower}.

 We are interested in obtaining bounds in terms of (generalizations of) quantum coherence. 
 %As we shall see, 
 There are different ways of defining quantum coherence in a quantum system. 
 First of all, there is the coherence of the state in a given basis $\mathcal B=\{|i\rangle\}$. {The norm of coherence for the state $\rho$ can be defined as the weight of the off diagonal elements in the basis $\mathcal B$, namely $\mathbb C(\rho):=\sum_{i\ne j} |\rho_{ij}|^2$}  \cite{plenio,zanardi2000entangling,zanardi2001entanglement,zanardicoh,styliaris2019coherence}. If we define the dephasing super-operator $\mathcal D(\cdot)=\sum_i \Pi_i \cdot \Pi_i$, with $\Pi_i=|i\rangle\langle i|$ are rank one projectors, then % obeying $\Pi_i \Pi_j=\delta_{ij} \Pi_i$ and $\sum_j \Pi_j=I$.
in terms of the dephasing superoperator, we have that 
\begin{equation}
\mathbb C(\rho)=\|\rho-\mathcal D(\rho)\|_F^2,
\label{eq:deco1}
\end{equation}
where $\|A\|_F = \sqrt{\tr{A A^\dag}}$ is the {Frobenius norm}. A simple calculation shows that $\mathbb C(\rho)= \tr\rho^2- \tr(\mathcal D\rho)^2$. As quantum coherence in a state is a basis dependent notion, it is important to determine what is the relevant basis. In a typical quantum experiment, quantum coherence is relevant in the basis of observables that will display interference.

Here, we consider the instantaneous eigenbasis  
%of the eigenstates 
of the unitary evolution operator $U_t$:
\begin{equation}\label{unitarydec}
    U_t=\sum_{j=1}^n e^{i \theta_j(t)} \Pi_j(t),
\end{equation}
where $\rho_t = U_t \rho_0 U_t^\dag$.
For instance, if one deals with a time-independent Hamiltonian $H=\sum_{i} \lambda_i \Pi_i$, then $U_t=\sum_{i} e^{-i \lambda_i t} \Pi_i$. 

One can easily see that the work, expressed as the energy storage, can be written as \cite{Caravelli1}
\begin{eqnarray}
\label{eq:work0}
    W\ =W_A&=&\underbrace{\text{Tr}(U_t^\dagger H_0 [\rho_0,U_t])}_{A}, \\
        W\ =W_B&=&\underbrace{\text{Tr}(U_t[U_t^\dagger H_0,\rho_0])}_{B},\\
   W\ = W_C&=&\underbrace{\text{Tr}(\rho_0 [U_t,U_t^\dagger H_0])}_{C}.
\end{eqnarray}
Lower and upper bounds to work can be found by applying the Von Neumann's trace inequality to the energy storage, resulting in bounds in terms of the singular values of the Hamiltonian. While these bounds are given for completeness in App.~\ref{sec:appvn}, the present paper focuses on bounds in terms of the norm of the dephased operator.

The identities above show that the amount of work one can extract from a quantum battery is related to how much the initial state $\rho_0$ anticommutes with the evolution operator $U_t$ (A), with the operator $ U_t^\dagger H_0$ (B), or how much these last two operators anticommute with each other (C). 
The amount of non-commutativity between two operators is related to how much off diagonal weight one operator has in the eigenbasis of the other. For a state, this is the definition of quantum coherence. More generally, we can define an operator-coherence in the basis $\mathcal{B}$ as
\begin{equation}
\mathbb C_{{\mathcal B}} (\hat{X}) := \frac{1}{2} \sum_j \|[\hat{X},\Pi_j]\|^2_F.
\label{eq:deco2}
\end{equation}
We refer to the above super operator $\mathbb C(\cdot)$ as \textit{generalized coherence operator} and the quantity $\mathbb C _{{\mathcal B}}(\hat{X})$ as the generalized coherence of {an arbitrary operator} $\hat{X}$ in $\mathcal B$.
%\LP{it's a number, so shall we change to ``generalized operator coherence'' or ``generalized coherence''?}. {\bf AH YES!} {Bin: I guess by operator, Francesco means the operator $C()$, though it's a super operator. I kept the ``super operator $\mathbb C(\cdot)$'' here for consistency -- in the following there are many places Francesco called $C()$ and operator, and also in the following the quantity $C_A$ is explicitly named the generalized operator coherence.}
In the following, all the definitions of norms are provided in App.~\ref{sec:appdef}.
The proof of the identity relationship between Eq.~(\ref{eq:deco2}) and Eq.~(\ref{eq:deco1}) is provided for completeness in App.~\ref{sec:appboundsec}.
For quantum states, this reduces to the usual definition of Frobenius-norm coherence. 
Given the notation above, let us now introduce some useful bounds that we will use in the following, and proved in App.~\ref{sec:appbounds}. {We will denote with $\|\cdot \|$ the operator norm
%\textcolor{blue}{, with $\|\cdot\|_2$ the 2-norm}
and with $\|\cdot \|_F$ the Frobenius norm.} 

In general, one has the following Frobenius inequality:
\begin{eqnarray}
|\text{tr}(A^\dagger B)|\leq \|A\|_F \|B\|_F.
\label{eq:traceineq}
\end{eqnarray}
Also, following from H\"{o}lder inequality,  one has (see App.~\ref{sec:appdef} for details):
\begin{eqnarray}
|\text{tr}(A^\dagger B)|\leq \|A\|\sqrt{\text{r}(B)} \|B\|_F,
\label{eq:rankbound}
\end{eqnarray}
%\LP{should it be $r(A)$ instead in the sqrt?}
where $\|A\|$ is the operator norm, that is, the maximum eigenvalue of the operator $A$; and $\text{r}(B)$ is the rank of the operator $B$, i.e., the number of its non-zero eigenvalues.
Moreover, we will use the following {two} inequalities:

\textbf{Lemma 1  - Single Normal Coherence Inequality.}
Let $A$ be a normal operator and let $B$ be an Hermitian operator. 
Then,
\begin{eqnarray}
\|[A,B]\|_F^2  &\leq&4 \|A\|^2 \mathbb C_A(B).
\label{eq:singleuni}
\end{eqnarray}

{\textit{Proof.}
Let $U=\sum_{i} \eta_i \Pi_i$, with $\Pi_{i}\Pi_j=\delta_{ij} \Pi_i$. Then,
we have 
\begin{eqnarray}
\|[U,A]\|_F^2&=& \sum_{ij} \eta_i \eta_j^* \text{tr}([\Pi_j,A]^\dagger [\Pi_i,A]) \nonumber \\
&=&2\sum_{ij} \eta_i \eta_j^* \text{tr}\Big(A^2\Pi_i \delta_{ij}-A \Pi_i A \Pi_j \Big)\nonumber \\
&=&2\sum_{i} |\eta_i|^2 \text{tr}\Big(A^2\Pi_i-A \Pi_i A \Pi_i \Big)\nonumber \\
&-&2\sum_{i\neq j} \eta_i \eta_j^* \text{tr}\Big(A \Pi_i A \Pi_j \Big)
\end{eqnarray}
We note that $2\sum_{i} |\eta_i|^2 \text{tr}\Big(A^2\Pi_i-A \Pi_i A \Pi_i \Big)\leq 2 \text{sup}_i |\eta_i^2 \mathbb{C}_{U}(A)$ and that
\begin{eqnarray}
-2\sum_{i\neq j} \eta_i \eta_j^* \text{tr}\Big(A \Pi_i A \Pi_j \Big)&\leq& 2 \sum_{ij} |\eta_i \eta_j^*|  \ |a_{ij}|^2\nonumber \\
&\leq&  2 |\eta_i^2 \mathbb{C}_U(A).
\end{eqnarray}
It follows that
$$ \|[U,A]\|_F^2  \leq4 \| A\|^2\mathbb{C}_U (A), $$
as claimed. $\square$}

{A similar lemma applies to the following commutator,
 whose proof is shown in App.~\ref{sec:appbounds}.}
 %being very similar to Lemma 1.}

\textbf{Lemma 2 - Double Normal Coherence Inequality.}  Assume A to be a normal operator and let B be Hermitian. Then, {the following bound applies}:
\begin{equation}
\|[A^\dagger,BA]\|_F^2\leq 4\| A\|^4 \mathbb{C}_A(B).
\label{eq:doubleuni}
\end{equation}
The bound above applies in the case of open systems as we will see soon.
Also, according to Proposition 2 in App.~\ref{sec:appbounds},
one has $\|[A^\dagger,BA]\|_F^2=\|[A^\dagger,AB]\|_F^2$. In the following we will use also the subadditivity property of the rank, e.g. given two operators $A:\mathbb{R}^n\rightarrow \mathbb{R}^n $, $B:\mathbb{R}^n\rightarrow \mathbb{R}^n $, we have $\text{r}(A+B)\leq \text{r}(A)+\text{r}(B)$. Clearly, however, $\text{r}(A+B)\leq \text{min}\Big(\text{r}(A)+\text{r}(B),n\Big)$, and if $U$ is a unitary operator, $\text{r}(U^\dagger A U)=\text{r}(A)$. Also, as a side remark, we note that by using a different procedure, also bounds in terms of $l_1$ coherence can be obtained (see for instance Lemma 2(b) in App.~\ref{sec:appbounds}).

\subsection{Bounds in terms of the generalized coherence}\label{sec:coh1}
 Given the definitions above,
we prove  the following three upper bounds:
\begin{eqnarray}\nonumber
  |W(t)|&\underbrace{\leq}_A& 2  \| H_0\| \sqrt{ \text{min}\Big(2\text{r}(\rho_0),n\Big) \mathbb C_{U_t}(\rho_0)}\\
   \nonumber 
   &\underbrace{\leq}_B&  2  \|\rho_0\|\sqrt{ \mathbb C_{\rho_0}(U_t^\dagger H_0)}\\
  &\underbrace{\leq}_C&  2 \|\rho_0\|_F   \sqrt{ \mathbb{C}_{U_t} (H_0)}
  \label{eq:goodbadugly}
\end{eqnarray}
with $n$ the dimension of the Hilbert space. We also assumed that $\rho_0:\mathbb{R}^n\rightarrow \mathbb{R}^n$. 
The expressions (A) and (C) bound the energy storage in terms of the quantum coherence of $\rho_0$ and $H_0$ in the basis of $U$.  One problem with the above bounds is that both the rank of $H_0$ or the $2-$norm  operator coherence $\mathbb C (X) $ may scale with the dimension $n$ of the Hilbert space. In this case, these bounds can become very loose for a high-dimensional systems.  On the other hand, for low-dimensional systems, or for large $n$ systems with low coherence, they are tighter and turn out to be useful.

We study specific models in Sec.~\ref{sec:models}. The bound (B) relates work to the purity of the initial state, as $\|\rho\|_F^2= \tr(\rho^2)$. For a pure initial state work is upper bounded by the operator norm. 
%, which is obvious.
The bound is saturated when the system is prepared in the highest excited state. One can obtain a further bound by substituting the operator norm of $\rho$ with its purity. 
%A bound similar to (B){ has been found in \cite{check}. }
As a side comment, it is interesting to note that since the bounds (\ref{eq:goodbadugly}) can be also obtained in the interaction representation, this is also true if we use $\rho_I$ in the interacting representation $\rho_I=U_0^\dagger \rho U_0$, $U_0=e^{-i H_0 t}$ and $\Pi_j$'s are in the interaction representation as well.

{\em Inequality A}.--- We start with the inequality $|\text{Tr}(AB)|\leq \sqrt{\text{r}(B)} \|A\| \|B\|_F,$ where $\text{r}(X)$ is the matrix rank (see App.~\ref{sec:appdef} for a proof).
We now pick $A=H_0$ and $B=U_t^\dagger [\rho_0,U_t]$. We have $\|U_t^\dagger [\rho_0,U_t]\|_F\leq \|[\rho_0,U_t]\|_F$. Now note that $\text{r}(U_t^\dagger [\rho_0,U_t])\leq \text{min}(2 \text{r}(\rho_0),n)$ because of the subadditivity of the rank.

If $\rho_0$ has a smaller rank, e.g., $\text{r}(\rho_0)< n/2$, then using Eq.~(\ref{eq:singleuni}), we obtain
\begin{eqnarray}
|W_A|&\leq& \|H_0\|\sqrt{2 \text{r}(\rho_0) \|[\rho_0,U_t]\|_F^2} \nonumber \\
%&=&\|H_0\| \sqrt{\|[\rho,U_t]\|_F^2 =\|H_0\|\sqrt{ \|[\rho_0,U_t]\|^2_F}\\
&=& \|H_0\|\sqrt{2 \text{r}(\rho_0) 4 \mathbb C_{U_t}(\rho_0)} \nonumber \\
&=&2 \sqrt{2} \| H_0\| \sqrt{ \text{r}(\rho_0) \mathbb C_{U_t}(\rho_0)},
\end{eqnarray}
which is the upper bound we reported above.

{\em Inequality B}.--- If we instead start with the general bound \cite{frobnorm,frobnorm2}
\begin{equation}
    \|[A,B]\|_F^2 \leq 2\|A\|_F  \|B\|_F -2 (\tr{A^\dagger B})^2,
\end{equation}
the bound would be loose, as $\|H_0\|_F$ is generally very large. We can then apply the trace inequality of Eq.~(\ref{eq:traceineq}), and obtain the following upper bound:
\begin{equation}
|W_B||\text{Tr}(U_t[U_t^\dagger H_0,\rho_0])| \leq \|U_t\|_F  \sqrt{ \|[U_t^\dagger H_0,\rho_0])\|_F^2}.
\end{equation}
We can now use $\rho_0$ as the operator in which we perform the spectral basis expansion. Then, using Lemma 1, Eq.~(\ref{eq:singleuni}), we have
\begin{equation}
|\text{Tr}(U_t[U_t^\dagger H_0,\rho_0])| \leq 2\|\rho_0\|  \sqrt{ \mathbb C_{\rho_0}(U_t^\dagger H_0)}
\end{equation}
as claimed.

{\em Inequality C}.--- We start with the trace inequality of Eq.~(\ref{eq:traceineq}), with $|\text{Tr}(AB)|\leq \|A\|_F \|B\|_F$. Using the fact that $\sqrt{\text{tr}(\rho^2)}=\|\rho\|_F=\sqrt{\text{purity}(\rho)}\leq 1$, we obtain
\begin{equation}
    |W_C|=\text{Tr}(\rho_0[U_t,U_t^\dagger H_0])\leq \|\rho_0\|_F \|[U_t, U_t^\dagger H_0]\|_F.
\end{equation}
Let us now focus on $\|[U_t, U_t^\dagger H_0]\|_F^2 $. Using the spectral decomposition for the unitary operator $U_t$, we have
\begin{eqnarray}
     \\|[U^\dagger,H_0A]\|_F^2\leq 4\| A\|^4 \mathbb{C}_A(B).
\end{eqnarray}
We see that the expression above is not in the form which can be directly expressed in terms of coherence.
Given the inequality above, then we have the following  inequality 
\begin{eqnarray}
    |W_C|&\leq& 2 \|\rho_0\|_F   \sqrt{ \mathbb{C}_U(H_0)},
\end{eqnarray}
which is the inequality we reported above. This bound has an interesting application in terms of designing better quantum batteries, as it states that better energy transfer can be achieved by a) starting from a pure state and b) using a driving system that is maximally not commuting with the Hamiltonian $H_0$ defining the energy. 

%For certain many-body systems, these bounds may be far from being saturated. %In particular, in cases when $X$ does commute with $$
%A direct bound from Eq.~\eqref{eq:work0} gives that $W \leq 2 \|H_0\|$, where the spectral norm is given by the largest absolute value for an operator's eigenvalues, $\|H\| = E_{max}$. Writing the Hamiltonian as $H = \sum_k E_k \ket{E_k} \bra{E_k}$,
%\begin{align}
%    W &= \tr{H_0 (\rho_0 - U_t \rho_0 U_t^\dag) } \\
%    &= \sum_k E_k \bra{E_k}(\rho_0- U_t \rho_0U_t^\dag) \ket{E_k} \\
%    &\leq E_{max} \sum_k ( \bra{E_k}\rho_0\ket{E_k} + \bra{E_k} U_t \rho_0U_t^\dag \ket{E_k} ) \\
%    &\leq 2 E_{max}.
%\end{align}
%Typically, the maximum energy $E_{max}$ of a many-body Hamiltonian grows with the number of constituent particles $N$. .

%

\subsection{Power and OTOC for closed systems}
Similar bounds to those we have obtained for the energy storage in the case of the closed system can be obtained for the power, that is, for the rate of change of the energy as a function of time. The energy can be written as 
\begin{eqnarray}
W(t)&=&W(0)+\int_0^t \frac{d}{dt'}W(t') dt' \nonumber \\
&=&   W(0)+\int_0^t P(t^\prime) dt^\prime \nonumber \\
&\leq& W(t)\ t\ \text{sup}_{t^\prime\in[0,t]} |P(t^\prime)|.
\end{eqnarray}
This inequality will be useful in later sections in particular in the case of open systems.

Bounds for the power can then be useful for obtaining bounds on the energy.  In particular, as we will see below, the bounds on the power can be still expressed in terms of the coherence operator $\mathbb C_Q(\cdot)$, but where as we will see in a moment $Q$ is different from the unitary operator.

%Given this preamble, 
Let us then consider the power in a closed system. The power is given by 
\begin{eqnarray}
    P_t&=&|\frac{d}{dt}W|=|\text{Tr}\Big(H_0 \dot \rho_t\Big)| \nonumber \\
    &=&|\text{Tr}\Big(H_0\Big[H_0+V,\rho_t\Big]\Big)|.
\end{eqnarray}
Following the three formulations of the work formula as in the previous sections, we have
\begin{eqnarray}
    P_t=\frac{d}{dt}W&=&\text{Tr}\Big(H_0 \dot \rho_t\Big) =\text{Tr}\Big(H_0\Big[H_0+V,\rho_t\Big]\Big) \nonumber \\
    &\underbrace{=}_A&\text{Tr}\Big(H_0\Big[V,\rho_t\Big]\Big) \nonumber \\
    &\underbrace{=}_B&\text{Tr}\Big(V\Big[\rho_t,H_0\Big]\Big) \nonumber \\
    &\underbrace{=}_C&\text{Tr}\Big(\rho_t\Big[H_0,V\Big]\Big) \nonumber 
    \label{eq:powerid},
\end{eqnarray}
where in the last equality we introduced the interaction picture operators $V_I=U_0^\dagger V U_0$ and $\rho_t^I=U_0^\dagger \rho_t U_0$. 

{\textbf{Proposition - } The power obeys the following upper bounds}:
\begin{eqnarray}
  P&\underbrace{\leq}_A& 2 \|H_0\|\cdot \|V\|\sqrt{ \text{r}([V,\rho_t]) \cdot \mathbb C_V(\rho_t)}, \label{eq:ineqA}\\
    P&\underbrace{\leq}_B& 2 \|H_0\|\cdot \|V\|\sqrt{\text{r}([\rho_t,H_0]) \cdot \mathbb C_{H_{0}}(\rho_t)},  \label{eq:ineqB}\\
  P&\underbrace{\leq}_C&2 \|\rho_0\| _F \|V\| \sqrt{ \mathbb C_{V} (H_0)},\
   \label{eq:ineqP}
\end{eqnarray}
where $ \epsilon$ and $\bar v$ are the maximum eigenvalues of the operators $H_0$ and $V$.

\textit{Proof.} 
 As for the case of the work, we see that the three different formulations of the work lead to three different inequalities. 
 We assume that $H_0$ and $V$ are both Hermitian, and that we can write both the Hamiltonian and the interaction operator in a spectral decomposition
\begin{equation}
    H_0=\sum_{j} \epsilon_j \Pi_j^{h},  \ \   V=\sum_{j} v_j \Pi_j^v,
\end{equation}
where $\Pi_j$'s are projector operators on the spectral basis of $H_0$ and $V$ respectively.  Let us focus on the first inequality. Using the decomposition above, 
\textit{inequality A} of Eq.~(\ref{eq:ineqA}) can be written as
\begin{eqnarray}
    P&\leq& \|H_0\|\sqrt{ \text{r}([V,\rho_t])\cdot  \|[V,\rho_t]\|_F^2 }.  \nonumber
\end{eqnarray}
We now use Lemma 1, Eq.~(\ref{eq:singleuni}), using as the normal decomposition operator $V$. We thus have 
\begin{equation}
    P\leq 2  \|H_0\|\cdot\|V\|\sqrt{ \text{r}([V,\rho_t])  \cdot \mathbb C_V(\rho_t)},  \\
\end{equation}
where $\mathbb C_V(\rho_t)$ is the coherence of $\rho_t$ in the  eigenbasis of the interaction operator $V$. Since $\rho_t$ has the same rank of $\rho_0$, and $\text{r}(AB)\leq\text{r}(A)$, we have $\text{ r}([V,\rho_t])\leq \text{r}(V)+\text{r}(\rho_0)$. Thus, if $V$ the interaction of a many body system it can be very large and thus reduce the applicability or tightness of this bound.

Let us now look at \textit{Inequality B}. We have
\begin{eqnarray}
P&\leq & \|V\|\sqrt{\text{r}([\rho_t,H_0])\cdot \|[\rho_t,H_0]\|_F^2}.
\end{eqnarray}
If we expand $H_0=\sum_{k} \epsilon_k \Pi_k^{h}$, then following the steps of the previous inequality we obtain 
\begin{eqnarray}
P&\leq & 2 \|H_0\|\ \|V\|\sqrt{\text{r}([\rho_t,H_0]) \cdot \mathbb C_{H_{0}}(\rho_t)},
\end{eqnarray}
where we see that also here we have $\text{r}([\rho_t,H_0]) \leq \text{r}(\rho_t)+\text{r}(H_0)$, and for a many body system such quantity can be very large.

At last, let us now focus on \textit{Inequality C} of Eq.~(\ref{eq:ineqB}). We can use in this case Eq.~(\ref{eq:traceineq}), and obtain
\begin{eqnarray}
     P&\leq& 2 \|\rho_t\|_F \sqrt{\| [H_0,V]\|_F^2 },
\end{eqnarray}
and use Lemma 1 again, in both the eigenbasis of $H_0$ and $V$. Note that since the evolution is unitary, we have $\|\rho_t\|_F=\|\rho_0\|_F$. We thus obtain two equivalent inequalities
\begin{eqnarray}
     P&\leq& 2 \|\rho_0\|_F \|V\| \sqrt{\mathbb C_{V}(H_0)} \nonumber \\
     &\leq& 2 \|H_0\| \|\rho_0\|_F \sqrt{\mathbb C_{H_0}(V)},
\end{eqnarray}
which shows that the the power depends also on the coherence of $H_0$ in the basis of $V$ and viceversa, and on the purity of the state. Clearly, this is a result of the fact that if the interaction $V$ commutes with $H_0$ no work can be done. Bound $C$ is the most interesting, as $\text{purity}(\rho_t)<1$ and the rank of $H_0$ or $V$ does not appear explicitly. This said, both ${\mathbb C}_{H_0}(V)$ or ${\mathbb C}_{V}(H_0)$  can be large. $\square$

 Interestingly, we note that the bounds for $P$ take the form of  Out-of-Time-Order-Correlators (OTOC)~\cite{isospec1,webqc,infoscrambling,Yan2020Recovery} at infinite temperature,
 which grow fast in time for typical $V$ and $\rho$.
 Moreover, note that inequalities A and B can be combined to obtain, 
 \begin{eqnarray}
 \left(\frac{P}{\|H_0\|\cdot \|V\|}\right)^2\leq \text{min}\big(\mathcal Q(\rho_t),\mathcal T(\rho_t)\big),
 \end{eqnarray}
 where we defined $\mathcal Q=\text{r}([\rho_t,H_0]) \|[\rho_t,H_0]\|_F^2$ and $\mathcal T=\text{r}([\rho_t,V]) \|[\rho_t,V]\|_F^2$.
Here, we see observe that if $\rho_t$ commutes with all  projectors, the bound collapses to zero. 
Finally, inequality C shows that the power is zero for $[H_0,V]=0$, as we discussed for the case of the total work. 
%Interestingly, both inequalities $A$ and $C$ are in the form of the OTOC.

\section{Coherence and energy in open quantum batteries}

In the previous sections we discussed the case of closed quantum batteries undergoing unitary evolution. More generally, we can consider the case of non-unitary quantum evolutions that describe, e.g., open quantum systems.  %We now focus on the case of open quantum systems, intended as a quantum battery coupled with a bath, and arrays of interacting batteries. 
%Since both fall within the same formalism as we will see, the bounds can be applied in similar ways.

\subsection{Bounds in terms of Kraus operators}
%We now consider the case of an open system approach.
%Intuitively, 
A generic quantum operation can be described by a quantum channel, that is, a completely positive, trace preserving map
\begin{equation}
    \mathcal E: \rho\mapsto \rho'=\mathcal E(\rho)
\end{equation}
In the Kraus operator sum representation such a map can be described as
\begin{equation}
    \mathcal E(\rho) = \sum_k A_k \rho A_k^\dagger
\end{equation}
where the Kraus operators $A_k$ are linear operators satisfying $\sum_{k} A_k^\dagger A_k=I$. This description can be used to model the evolution of a subsystem interacting with other quantum systems. Indeed, for a composite system $\mathcal H= \mathcal H_{\mathcal S}\otimes\mathcal H_B$ evolving unitarily with a Hamiltonian $H\in\mathcal B(\mathcal H)$ by $U=\exp(-iHt)$, the evolution of the reduced system is given by a completely positive map $\rho_{\mathcal S} =\tr_B(U\rho U^\dagger)\equiv  \sum_k A_k \rho_A A_k^\dagger$. 
For open quantum systems, or more generally for non unitary evolution, work cannot be identified with the difference of energy but one has to consider also variations of entropy. Here we focus thus just on the energy exchanged $\Delta E(t)$.

Now we move on to generalize the bounds to energy exchange in terms of coherence for the evolution map $\mathcal E_t$

%In order to perform the bounds we need to take into account the interaction with the bath. 
%From the point of view of the action of the bath on the battery, the evolution of density matrix for the battery subsystem is governed by the action of the POVM:
%\begin{eqnarray}
%\rho_{\mathcal S}(t)=\sum_a A_a(t)  \rho_{\mathcal S}(0)  A^\dagger_a(t).
%\end{eqnarray}
%We wish to generalize the results of the previous section on the closed battery to the case of the POVM evolution of the density matrix, and in particular in terms of the Kraus operators.
%With this, we can obtain similar upper bounds for the energy exchange $\Delta E$ as in the case of the closed system.
Let us define $\rho_a(t)= A_a \rho(0) A_a^\dagger$. Also, we define $\mathcal E$ as
\begin{eqnarray}
\mathcal E(X)=\sum_{a} A_a X A_a^\dagger,\ \ \ \ \mathcal E_{*}(X)=\sum_{a} A_a^\dagger X A_a.
\end{eqnarray}
First, we note that because of the Hilbert-Schmidt duality, we have
\begin{eqnarray}
\text{Tr}(\mathcal E(\rho) H_0)=\text{Tr}(\rho \mathcal E_{*}(H_0)).
\end{eqnarray}
Then, we can rewrite
\begin{eqnarray}
\Delta E&=&\sum_a \text{Tr}\Big(\big(\rho_{\mathcal S}(0)-\rho_a(t)\big)H_0\Big)\nonumber \\
&\underbrace{=}_A&\sum_a \text{Tr}\Big(A_a H_0[A_a^\dagger,\rho_{\mathcal S}(0)]\Big) \nonumber \\
&\underbrace{=}_B&\sum_a \text{Tr}\Big(A_a^\dagger[ \rho_{\mathcal S}(0),A_a H_0 ]\Big) \nonumber \\
&\underbrace{=}_C&\sum_a \text{Tr}\Big(\rho_{\mathcal S}(0)[  A_a H_0,  A_a^\dagger]\Big), 
\label{eq:equalities}
\end{eqnarray}
We can upper bound $|\Delta E|$ as in the case of the closed quantum system, with the difference that now instead of a unique evolution operator, we have a Kraus representation, which albeit satisfying $\|\sum_a A_a^\dagger  A_a \|=\| I\|=1$, are not unitary operators. 
We can also write 
$\| A_{j}^\dagger+\sum_{a\neq j} A_a^\dagger  A_a   A_{j}^\dagger\|=\|  A_{j}^\dagger\|$.

{\textbf{Proposition} 
%Given the equalities of eqn. (\ref{eq:equalities}) and provided that $A_a$ is Hermitian,the three following inequalities apply respectively:
The following inequalities apply for Hermitian $A_a$:
}
\begin{eqnarray}
|\Delta E|&\underbrace{\leq}_A&2  \|H_0\| \sum_{a} \| A_a\|^2 \sqrt{ R_a^1\mathbb{C}_{a}(\rho_{\mathcal S}(0))}, \nonumber \\
|\Delta E|&\underbrace{\leq}_B&2 \|\rho_{S}(0)\| \|H_0\|  \nonumber \\
& &\cdot  \sum_{a} \| A_a\| \sqrt{ R_a^2\ \mathbb{C}_{\rho_{\mathcal S}(0)}( A_a H_0)}, \nonumber  \\
|\Delta E| &\underbrace{\leq}_{C}&  2 \|\rho_{\mathcal S}(0)\|_F \sum_a \|A_a\|^2 \sqrt{{\mathbb C}_a(H_0)} ,
\end{eqnarray}
where $R_a^1=\text{r}([A_a^\dagger,\rho_{\mathcal S}])$ and $R_a^2=\text{r}([\rho_{\mathcal S}(0), A_a H_0])$; $\mathbb C_a(\cdot) $ is the coherence operator 
in the spectral resolution of $A_a$, while $\mathbb{C}_{\rho_{\mathcal S}(0)}(\cdot)$ is the generalized coherence operator in the spectral basis of the reduced density matrix.

\textit{Proof.} As an example, let us consider first the latter equality in Eqs.~(\ref{eq:equalities}). 
We can write, for \textit{Equality C},
\begin{eqnarray}
\Delta E&=&\sum_a\text{Tr}\Big(\rho_{\mathcal S}(0)  [A_aH_0,A_a^\dagger]\Big).\nonumber
% &=&\sum_a\text{Tr}\Big(\rho_{\mathcal S}(0)  [A_a H_0,A_a^\dagger]\Big)\nonumber
\end{eqnarray}
We can thus write,  using the trace inequality of Eq.~(\ref{eq:traceineq}), the following expression:
\begin{eqnarray}
|\Delta E| \leq  \|\rho_{\mathcal S}(0)\|_F \sum_a \cdot \sqrt{\| [A_a H_0,A_a^\dagger]\|_F^2}.
\end{eqnarray}

We now make some assumptions about the Kraus operators, i.e., that these are Hermitian operators, in order to obtain an upper bound for $\|[A_a H_0,A_a]\|_F$ in terms of the coherence of the Hamiltonian in the $A_a$ basis. While this assumption is unphysical, it is a loss of generality and extension of these bounds will be considered in the future. 

Using Lemma 2 in Eq.~(\ref{eq:doubleuni}), we have that if $A_a$ is Hermitian, then, 
\begin{eqnarray}
|\Delta E| \leq  2 \|\rho_{\mathcal S}(0)\|_F \sum_a \|A_a\|^2 \sqrt{{\mathbb C}_a(H_0)},
\end{eqnarray}
where $\mathbb C_{a}(\cdot)$ is the coherence operator in the spectral basis of $A_a$,
\begin{eqnarray}
    \mathbb C_a(H_0)=\frac{1}{2}\sum_{j} \|[H_0,\hat \Pi_a^j]\|_F^2.
\end{eqnarray}
%{check the proof detail}
The final upper bound above is given in terms of the sum of the coherences of $\rho_0$ with respect to the Kraus operators $A_a$, with the assumption that $\text{r}(H_0)\gg 1$. While these bounds might be useful in some cases, typically it is easier to work in terms of Lindblad operators, as we we do in the next section.

If instead we start from \textit{Equality A} and \textit{B} in Eqs.~(\ref{eq:equalities}), and using the same procedure we used in the closed system, we obtain the bounds thanks to Lemmas 1 and 2, we obtain the following bounds:
\begin{eqnarray}
\|[A_a^\dagger,\rho_{\mathcal S}]\|_F&\leq & 2 \|A_a\| \sqrt{\mathbb{C}_{a}(\rho_{\mathcal S}(0))}, \nonumber \\
\|[\rho_{\mathcal S}(0), A_a H_0]\|_F&\leq & 2  \|\rho_{S}(0)\|  \sqrt{\mathbb{C}_{\rho_{\mathcal S}(0)}( A_a H_0)}.\nonumber 
\end{eqnarray}
If we combine these with the inequality  of Eq.~(\ref{eq:rankbound}), we obtain
\begin{eqnarray}
|\Delta E|&\underbrace{\leq}_A&2  \|H_0\| \sum_{a} \| A_a\|^2 \sqrt{ R_a^1\mathbb{C}_{a}(\rho_{\mathcal S}(0))}, \nonumber \\
|\Delta E|&\underbrace{\leq}_B&2 \|\rho_{S}(0)\| \nonumber \\
& &\cdot\|H_0\| \sum_{a} \| A_a\| \sqrt{ R_a^2\ \mathbb{C}_{\rho_{\mathcal S}(0)}( A_a H_0)}, \nonumber
\end{eqnarray}
where $R_a^1=\text{r}([A_a^\dagger,\rho_{\mathcal S}])$ and $R_a^2=\text{r}([\rho_{\mathcal S}(0), A_a H_0]) $. $\square$

\subsection{Bounds in terms of Lindblad operators}\label{sec:lindblad}
Consider the case of a quantum system $\mathcal H_S$ interacting with a bath $\mathcal H_B$.
Under standard assumptions such as Markovian evolution and small couplings to the environment, the evolution of the density matrix $\rho_S$ of the subsystem can be described by the Liouvillian operator $\mathcal L$ through the master equation:

%Interesting bounds can be obtained in terms of Lindblad operators as follows.

%\subsection{Kraus vs Lindblad rotations}
%Let us now note the following relation between the Lindblad and Kraus operators for the case of Markovian evolution. If the dynamics is Markovian, we can write in terms of the Liouvillian operator $\mathcal L$ the following representation of the master equation:
\begin{eqnarray}
\frac{d}{dt} \rho_{\mathcal S}(t)= \mathcal L(\rho_{\mathcal S}(t)).\nonumber
\end{eqnarray}
We can then write
\begin{eqnarray}
\rho_{\mathcal S}(t)=\lim_{n\rightarrow \infty}(1+\frac{\mathcal L t}{n})
^n=e^{\mathcal L t} \rho_{\mathcal S}(0).\nonumber
\end{eqnarray}
Then if we assume that $\rho_{\mathcal S}(t)$ has a Kraus representation, at the order $dt$ we must have
\begin{eqnarray}
\hat A_0&=&I+dt(-i \hat H +\sum_{i=1} \hat L_i \hat L_i^{\dagger}),\nonumber \\
\hat A_i&=&\sqrt{dt}\ \hat L_i,i\geq 1,\nonumber
\end{eqnarray}
see, e.g., \cite{krauslindblad}.

The Kraus operators in fact contain both the battery Hamiltonian and the interaction with the bath.  These enter in a non-trivial way into the density matrix evolution for the battery in terms of Kraus operators. One way to overcome this problem is via the fundamental theorem of calculus, which gives
\begin{eqnarray}
E(t)-E(0)=\int^t_0 \frac{d E}{dt}  dt=\int^t_0 P_{t^\prime}dt^\prime.
\label{eq:enstor}
\end{eqnarray}
In order to bound the energy difference for $t>0$, we can use the integral inequality, e.g. that for arbitrary scalar functions $A(\tau)$ we have $\int_0^t A(\tau)d\tau\leq\int_0^t |A(\tau)|d\tau\leq t\ \text{sup}_{\tau\in[0,t]} |A(\tau)|$. This bound works in particular well for short times $t$.
In an open system we have that
\begin{eqnarray}
    \frac{d}{dt} E(t)&=&\text{tr}\left( \frac{d}{dt} \rho_{\mathcal S}(t) H_0 \right) \nonumber \\
    &=&\text{tr}\bigg(\Big(\frac{i}{\hbar } [\rho_{\mathcal S},H_0+V]+\mathbb{ L}(\rho_t)\Big)H_0\bigg),\nonumber 
\end{eqnarray}
where
\begin{eqnarray}
    \mathbb{ L}(\rho_{\mathcal S})= \sum_n\gamma_n\Big( L_n \rho_{\mathcal S}(t) L_n^\dagger -\frac{1}{2}\{L_n^\dagger L_n,\rho_{\mathcal S}(t)\}\Big)\nonumber 
\end{eqnarray}
with $\{\cdot,\cdot\}$ the anticommutators, and where the Lindblad operators $L_n$ satisfy $\text{tr}[L_p L_t^\dagger]=\delta_{pt}$.
We assume below that $L_{n}$ is Hermitian that, while not being the most general case, does include many physically relevant models. 

Following the definitions above, then, we prove the following 

{\textbf{Proposition} (Upper bound on stored energy).
Consider an open quantum system with Hamiltonian $H_0$ evolving under Hermitian Lindlad operators. 
%Let us consider an open quantum system composed of a bath, with its Hilbert space $\mathcal H_B$, and a system of interest $\mathcal H_s$, described by a Hamiltonian $H_0$. The action of the bath on the quantum system is described by Hermitian Lindblad operators. 
Then, the stored energy, as defined in eqn. (\ref{eq:enstor}), satisfies the following upper bound:
}
\begin{eqnarray}
    \frac{\hbar|E(t)-E_0|}{ \|H_0\|}\leq t(W_A +W_B ),
    \label{eq:opensysbound}
\end{eqnarray}
where the quantity $W_A$ can be expressed in the following forms:
\begin{eqnarray}
    W_A= 2\ \text{min}_{\tau\in [0,t]} \begin{cases}
    \|V\| \sqrt{\text{r}([\rho_{\mathcal S}(\tau),V])\mathbb C_{V}(\rho_{\mathcal S}(\tau))} \\
     \|\rho_{\mathcal S}(\tau)\|_F  \sqrt{  \mathbb C_{H_0}(V)}\\
    \|V\|   \sqrt{\text{r}([\rho_{\mathcal S},H_0]) \mathbb C_{H_0}(\rho_{\mathcal S}(\tau))} ,
    \end{cases}
\end{eqnarray}
while 
\begin{eqnarray}
    W_B&=&3\sum_n \gamma_n   \text{sup}_{\tau\in[0,t]} \sqrt{r_n \mathbb C_{L_n}\big(\rho_{\mathcal S}(\tau)\big) },  
    \end{eqnarray}
where  $C_{L_n}(\rho_{\mathcal S})$ is the coherence in the spectral basis of $L_n$ of $\rho_{\mathcal S}(\tau)$. Let us call $\Pi_i^n$ the spectral basis of $L_n$. Then,
\begin{eqnarray}
C_{L_n}(\rho_{\mathcal S})=\sum_{i} \|[\Pi_i^n,\rho_{\mathcal S}]\|_F^2.
\end{eqnarray}
The constant $r_n$ is given by $r_n=\text{r}(\rho_{\mathcal S})+\text{r}(L_n)$.

\textit{Proof.}
Following from the integral inequality, we begin with, 
\begin{eqnarray}
    \hbar |\Delta E(t)|&\leq& \int_0^t|\text{tr}([\rho_{\mathcal S}(\tau),V]H_0)|d\tau\nonumber \\
    &+&\int_0^t|\text{tr}(\mathbb{L}\big(\rho_{\mathcal S}(\tau)\big)H_0)|d\tau.
    \label{eq:powerb}
\end{eqnarray}
The first term can be bounded as we did in the case of the closed system, thus we have the identities
\begin{eqnarray}
    \underbrace{|\text{tr}([\rho_{\mathcal S}(\tau),V]H_0)|}_{W_A^1}&=&\underbrace{|\text{tr}([V,H_0]\rho_{\mathcal S}(\tau))|}_{W_A^2}\nonumber \\
    &=&\underbrace{|\text{tr}([\rho_{\mathcal S},H_0]V)|}_{W_A^3}.\nonumber 
\end{eqnarray}
It follows that can use, following the bounds of Sec.~\ref{sec:coh1}, we immediately write the three possible inequalities
\begin{eqnarray}
    W_A^1&\leq& 2 \|H_0\| \|V\| \sqrt{\text{r}([\rho_{\mathcal S}(\tau),V])\mathbb C_{V}(\rho_0)},  \nonumber \\
    W_A^2&\leq& 2 \|\rho_{\mathcal S}\|_F \|H_0\| \sqrt{  \mathbb C_{H_0}(V)}, \nonumber \\
    W_A^3&\leq&  2 \|V\| \|H_0\| \sqrt{\text{r}([\rho_{\mathcal S},H_0]) \mathbb C_{H_0}(\rho_0)} .
\end{eqnarray}
In the expressions above we defined $\mathbb C_{V}(\cdot )=\sum_k \|[\cdot ,\Pi_k^V]\|_F^2$,, and we wrote $\Pi_k^V$ is the spectral basis of the interaction operator $V$, and of course similarly $\mathbb{C}_{H_0}(\cdot)=\sum_k \|[\cdot ,\Pi_k^{H_0}]\|_F^2$, with $\Pi_k^{H_0}$ the spectral basis of the battery Hamiltonian.
Let us now focus on the Lindbladian. First, note that we have, because $\{A,B\}=[A,B]+2 BA$, 
\begin{eqnarray}
\mathbb{L}\big(\rho_{\mathcal S}(\tau)\big)&=&\sum_n \gamma_n\big( L_n \rho_{\mathcal S}L_n^\dagger-\frac{1}{2}\{L_n^\dagger L_n,\rho_{\mathcal S}\}\big) \nonumber \\
%&=&\sum_{n}\gamma_n\Big( L_n \rho_{\mathcal S} L_n^\dagger -\frac{1}{2} [L_n^\dagger L_n,\rho_{\mathcal S}]-\rho_{\mathcal S}L_n^\dagger L_n \Big)\nonumber \\
&=&-\sum_n \gamma_n\Big([\rho_{\mathcal S} L_n^\dagger,L_n]+\frac{1}{2}[L_n^\dagger L_n,\rho_{\mathcal S}(\tau)] \Big).\nonumber \\
\end{eqnarray}
We thus have
\begin{eqnarray}
    \hbar |\Delta E(t)|&\leq&W_A\nonumber \\
        \label{eq:firstterm}
    &+&\sum_n\frac{\gamma_n}{2}\int_0^t  |\text{tr}\Big([L_n^\dagger L_n,\rho_{\mathcal S}(\tau)]H_0\Big)| d\tau \nonumber \\
    &+&\sum_n \gamma_n\int_0^t  |\text{tr}\Big([\rho_{\mathcal S}(\tau)L_n^\dagger  ,L_n]H_0\Big)| d\tau.\nonumber \\
    \label{eq:sumopen}
\end{eqnarray}
Let us now make the assumption  the Lindblads are Hermitian operators, e.g. $L_{n}^\dagger=L_n$. This bound is the one we use in the following for the case of the Spin-Boson model, for which such condition holds true. %Moreover, we use the fact that $\text{tr}\Big([\rho_{\mathcal S}(\tau) L_n ,L_n]H_0\Big)=\text{tr}\Big([ \rho_{\mathcal S}(\tau) [ L_n,L_n H_0]\Big)$. 

For the term of Eq.~(\ref{eq:firstterm}), we use the trace inequality of Eq.~(\ref{eq:rankbound}), and the fact that  $L_n^\dagger L_n$ is an Hermitian operator, it has a spectral decomposition of the form
\begin{eqnarray}
    \tilde L_n^2=L_n^\dagger L_n=\sum_{k} |\eta^n_k|^2 \Pi_k^n.
\end{eqnarray}
We can now use the bounds we used in terms of coherence. We have, using Lemma 1 and Eq.~(\ref{eq:singleuni}), we have
\begin{eqnarray}
    \|[L_n^\dagger L_n,\rho_{\mathcal S}(\tau)]\|_F^2\leq 4 \|L_n\|^4 \mathbb C_{L_n}\big(\rho_{\mathcal S}(\tau)\big),
\end{eqnarray}
where $\mathbb C_{L_n}\big(\rho_{\mathcal S}(\tau)\big)=\sum_k \|[\Pi_k^n,\rho_{\mathcal S}(\tau)]\|_F^2$. Since $\text{tr}(L_n^\dagger L_n)=1$, we have $\|L_n\|_F=1$ and $\|L_n\|\leq1$.

If we call $R_n^\prime=\text{r}([L_n^\dagger L_n,\rho_{\mathcal S}(\tau)])$, then
\begin{eqnarray}
\int_0^t& & |\text{tr}\Big([L_n^\dagger L_n,\rho_{\mathcal S}(\tau)]H_0\Big)| d\tau \nonumber \\
&\leq&2\ t\ \|H_0\| \cdot \text{sup}_{\tau \in [0,t]}  \sqrt{ R_n^\prime \mathbb C_{L_n}\big(\rho_{\mathcal S}(\tau)\big)}\nonumber
\end{eqnarray}

For the second term, in Eq.~(\ref{eq:sumopen}), we need to use the fact that $L_n$ is Hermitian.

If $L_n$ is an Hermitian operator, then it does have a spectral decomposition of the form $L_n=\sum_i \eta_i^n \Pi_k^n $, and it follows trivially that $\tilde L_n^2=\sum_i |\eta_i^n|^2 \Pi_k^n$. In this case, we obtain directly from Eq.~(\ref{eq:sumopen}) that,  calling $R_n^{\prime \prime}=\text{r}([\rho_{\mathcal S}(\tau) L_n^\dagger ,L_n])$, that
\begin{eqnarray}
 & &   |\text{tr}\Big([\rho_{\mathcal S}(\tau) L_n^\dagger ,L_n]H_0\Big)| \nonumber \\
    & &\ \ \ \ \ \ \ \ \ \ \leq\|H_0\|\ \sqrt{R_n^{\prime \prime}} \|[\rho_{\mathcal S}(\tau) L_n^\dagger ,L_n]\|_F.\nonumber 
\end{eqnarray}

At this point, we can use the fact that $L_n$ is Hermitian. We expand $L_n$ in the $\Pi$'s basis, obtaining from Lemma 2  of Eq.~(\ref{eq:doubleuni}):
$$\|[A,BA]\|_F^2\leq 4\|A\|^4 \mathbb C_{A}(B)$$.
Using the bound above, we have
\begin{eqnarray}
 \|[\rho_{\mathcal S}(\tau) L_n^\dagger ,L_n]\|_F&\leq& 2 \|L_n\|^2 \sqrt{\mathbb C_{L_n}(\rho_{\mathcal S}(\tau))}.\nonumber 
\end{eqnarray}
Thus,
\begin{eqnarray}
 \|[\rho_{\mathcal S}(\tau) L_n^\dagger ,L_n]\|_F&\leq& 2 \sqrt{\mathbb C_{L_n}(\rho_{\mathcal S}(\tau))}.
\end{eqnarray}

We now note that both $R_n^\prime\leq \text{r}(L_n)+\text{r}(\rho_{\mathcal S})$, and $R_n^{\prime \prime}\leq \text{r}(L_n)+\text{r}(\rho_{\mathcal S})$. Let us thus call $r_n=\text{r}(L_n)+\text{r}(\rho_{\mathcal S})$.  

Thus, we have,
\begin{eqnarray}
    \frac{\hbar |\Delta E(t)|}{\|H_0\|t}&\leq& W_A +3\sum_n \gamma_n   \text{sup}_{\tau} \sqrt{r_n \mathbb C_{L_n}\big(\rho_{\mathcal S}(\tau)\big) },  \nonumber \\
             \label{eq:gron}
\end{eqnarray}
which is the final result which we use in the statement above. $\square$

In the following, we will apply the bounds for the case of the slave boson model.

\subsection{Ensemble of batteries}
A setup of interest for practical applications is the case of an array of batteries interacting through a potential $V$, and we want to show how the coherence enters in this case. Here we assume that the system is closed, but the formalism is analogous to the case of the open system as we discuss below. We consider $M$ copies of the same system, a battery, coupled only via an interaction potential. Formally, such situation can be represented as follows. The $M$ copies of the same system are represented by a Hamiltonian of the form $H=\bigoplus_{i=1}^M H_i=\bigoplus_{i=1}^M (H_0^i+V^i)$, e.g. an ensemble of batteries If we assume that the $M$ copies are not interacting,  since $\rho(t)=\otimes_{i=1}^M\rho_i(t)$, we have
\begin{eqnarray}
W(t)&=&\sum_{i=1}^M \text{Tr}_i\Big(\big(\rho^i(t)-\rho^i(0)\big) H_0^i\Big)-E_0\nonumber \\
&\leq& M W_{max},\nonumber
\end{eqnarray}
where $W_{max}$ is maximum work obtained for the single system $H_0^i+V^i$. In this case we can apply the formalism of the closed system for each subsystem and thus it is not so interesting.
If however we only have that $H_0=\bigoplus_{i=1}^M H_0^i=\sum_{i=1}^M H_0^i\otimes I_{\mathcal S\setminus i}$, while $V$ does  have support on multiple batteries single battery, then we have that the density matrix is in general not factorizable, and the energy storage can be written in the form
\begin{eqnarray}
W(t)&=&\text{Tr}\big(\rho(t) (\sum_{i=1}^M H_0^i\otimes  I_{\mathcal S\setminus _i})\big)-E_0\nonumber \\
&=&\sum_{i=1}^M\text{Tr}_i\Big(\text{Tr}_{\mathcal S \setminus i}\big(\rho(t)\big) H_0^i\Big)-E_0,\nonumber 
\end{eqnarray}
{where we denote with $\mathcal S \setminus i$ as the system $\mathcal S$ without the \textit{battery} subsystem $i$, and}
where $\rho_i(t)=\text{Tr}_{\mathcal S \setminus i}\big(\rho(t)\big)$ is the local density matrix relative to subsystem $i$.
In this case, the time evolution of the reduced density matrix can be written in terms of a positive map \cite{petruccione,schlosshauer}. We can write $\rho_i(t)=\sum_a A_a^i(t) \rho_i(0) A_a^i(t)$, with $\sum_a A_a^i(t) (A_a^i(t))^\dagger=I$ where $A_a^i$ are Kraus operators. It follows that we can write
\begin{eqnarray}
W(t)&=&\sum_{i=1}^M\sum_{a}\text{Tr}_i\Big( \big(A_a^i(t) \rho_i(0) A_a^i(t)^\dagger-\rho_i(0)\big)  H_0^i\Big)\nonumber \\
&=&\sum_{i=1}^M\sum_{a}\text{Tr}_i\Big( \rho_i(0) [A_a^i, H_0^i(A_a^i)^\dagger]\Big),\nonumber 
\label{eq:open1}
\end{eqnarray}
where in the second line we have used the Hilbert-Schmidt duality. Here, we called $A_a^i$ the Kraus operators to avoid confusion with the open system case, but the meaning is similar.

If the Kraus operators $(A^i)^\dagger_a=A^i_a$ are Hermitian, similarly to what we have done before, they can be expanded via a spectral decomposition of the form $A_a^i=\sum_{k} (\eta_i^a)_k (\Pi_i^a)_k$, and we can apply Lemma 2 of Eq.~(\ref{eq:doubleuni}), proven in App.~\ref{sec:appbounds}.  Using the spectral basis decomposition,  then we have
\begin{eqnarray}
 \|[A_a^i,H_0^i A_a^i]\|_F^2&\leq& 4 \|A_a^i\|^4 \mathbb{C}_{ai}(H_0^i),
\end{eqnarray}
where $\mathbb{C}_{ai}(H_0^i)=1/2 \sum_{k} \|[(\Pi_i^a)_k,H_0^i]\|_F^2$.
We now use the trace inequality of Eq.~(\ref{eq:traceineq}), obtaining
\begin{eqnarray}
W(t)&\leq& \sum_{i=1}^M\sum_{a}\|\rho_i(0)\|_F \|[A_a^i, H_0^i(A_a^i)^\dagger]\|_F.\nonumber  \\
&\leq& 2 \sum_{i=1}^M\sum_{a}\|\rho_i(0)\|_F \|A_a^i\|^2 \sqrt{\mathbb{C}_{ai}(H_0^i)},
\end{eqnarray}
from which it follows that the energy storage of the system can be reduced to coherence bounds on the single subsystem.

\section{Models study}\label{sec:models}
In this section we study the work and charging power for some models. We consider two closed quantum systems. The first one is a two-body spin model, the second one a quantum  many-body system described by a  the anisotropic XY spin chain. The third model studied is the open quantum system of a single spin in a bosonic bath. %{Bin: Since a third Spin-Boson model was added, need to modify the above introduction.}

\subsection{A 2-body system example}

As described previously, the bounds we have obtained can be tight for small systems.
Here we consider first a 4-level system given by two spins $s_1$ and $s_2$ interacting with two external field $\vec G(t)$ and $\vec F(t)$ and among themselves \cite{Bragov1,Bragov2}.
The Hamiltonian of the system is assumed to be $H=H_0+V$, with $H_0=2 J \hat s_1\otimes \hat s_2$, where $\hat s_1= \vec \sigma\otimes I$ and $\hat s_2=I\otimes \frac{1}{2} \vec \sigma$. The external fields enter in $V$, with $V=2 \big( \hat h_1\otimes I+I\otimes \hat h_2\big)$, with $\hat h_1=\vec \sigma \cdot \vec G$, $\hat h_2=\vec \sigma \cdot \vec F$.

For simplicity, here we consider $\vec F$ and $\vec G$ to be aligned along the $z$ direction, thus $\vec G=(0,0,B_1)$ and $\vec F=(0,0,B_2)$, which is an exactly solvable model. In this system, the energy can be stored due to the coupling between the two spins. If one introduces $B_{\pm}=B_1(t) \pm B_2(t)$,the spinor $\Psi=(v_1,v_2,v_3,v_4)$ satisfies the equation
\begin{equation}
    i \partial_t \Psi=\hat H\Psi,\nonumber 
\end{equation}
with
\begin{eqnarray}
H&=&
{\footnotesize \begin{pmatrix}
B_{+}+J/2 & 0  & 0 & 0\\
0 & B_{-}-J/2 & J & 0 \\
0 & J & B_{-}-J/2 & 0 \\
0 & 0& 0 & J/2- B_{+}
\end{pmatrix}}\nonumber \\
&=&H_0+V,
\end{eqnarray}
where we assume $H_0$ to be dependent only on the $J$ coupling.
The components $1$ and $4$ of the spinor satisfy the solution, assuming that $J$ is constant,
\begin{eqnarray}
v_1(t)&=& e^{-i \int_0^t B_+(t^\prime) dt^\prime -i \frac{J}{2} t} v_1(0),\\
v_4(t)&=& e^{+i \int_0^t B_+(t^\prime) dt^\prime -i \frac{J}{2} t} v_4(0).
\label{eq:sol1}
\end{eqnarray}
If we define $\psi^\prime=(v_2,v_3)$, then these components satisfy the equation
\begin{equation}
    i\partial_t \psi^\prime(t)=(\vec \sigma \cdot \vec K-\frac{J}{2}) \tilde \psi^\prime(t),
\end{equation}
where $\vec K(t)=\Big(J,0,B_{-}(t)\Big)$.
\begin{figure}
    \centering
    \includegraphics[scale=1.5]{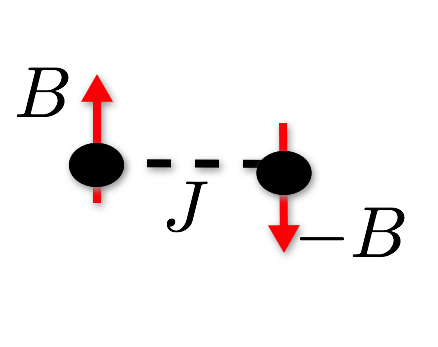}
    \caption{The second protocol proposed for the closed 4-level system via local fields with opposite orientation.}
    \label{fig:figuresecond}
\end{figure}
\subsubsection{Work and Power}

Let us thus consider the following two quench protocols. 

\textit{First protocol.} In the first, we have $B_1(t)=B_2(t)=\theta(t-t_0) B$. Then, $B_+=2 B \theta(t-t_0)$ and $B_-(t)=0$. Thus, the spin components $v_2$ and $v_3$ decouple from the external field. In this case, $\psi_a$ is a equivalent to a single spin interacting via a $\sigma_z$ Zeeman coupling with external field $B_+$. The work is given by $W(t)=\text{Tr}(\rho(t) H_0)-\text{Tr}(\rho_0 H_0)$, and the components $2$ are $3$ decoupled and do not change their populations. Thus, this protocol while natural is not particularly interesting. Using Eq.~(\ref{eq:sol1}) it is not hard to see that $W(t)=0$. Another way of seeing this is through the inequality $B$ of Eq.~(\ref{eq:goodbadugly}). We have in fact that $\Pi_V$'s are identical to the projectors on subspaces spanned by $H_0$, and thus $\mathbb C_{U_t}(H_0)=0$.

\textit{Second protocol.} We consider $B_1(t)=-B_2(t)=\theta(t-t_0) B$, as shown in Fig.~\ref{fig:figuresecond}. Then, $B_-=2 B \theta(t-t_0)$ and $B_+(t)=0$. Also in this case the components $v_1$ and $v_4$ decoupled and do not contribute to the change in energy. We follow all the steps in order to calculate the bounds later. The Stone operator for the reduced system is given by 
\begin{eqnarray}
U(t>t_0)&=&e^{-i \int_{t_0}^t(\vec \sigma \cdot \vec K(t)-\frac{J}{2}) dt }\nonumber \\
&=&e^{-i \Big(\sigma_x J (t-t_0)+2 B (t-t_0)\sigma_z-\frac{J}{2} (t-t_0)\Big). }\nonumber 
\end{eqnarray}
If we use the formula $e^{i a \hat n\cdot \vec \sigma}=I\cos(a)+i \sin(a) \hat n\cdot \vec \sigma$, then we can write, using $a=\sqrt{J^2+4 B^2}$ and $\hat n=(\frac{J}{a},0,\frac{2B}{a})$
\begin{eqnarray}
U(t>t_0)&=&e^{i \frac{J}{2} (t-t_0)}\big(I \cos(a(t-t_0)) \nonumber \\
& &\ \ \ +i \sin(a(t-t_0)) (\frac{J}{a}\hat \sigma_x+\frac{2B}{a} \hat \sigma_z)\big).\nonumber 
\end{eqnarray}
If we define $\rho_0$ on the basis of $v_2,v_3$, we have a base Hamiltonian for the battery defined as $H_0=J\sigma_x-\frac{J}{2} I$ (with eigenvalues $J/2,-3/2J$) and $V=2 B \hat \sigma_z$ (with eigenvalues $\pm 2 B$). We thus have $\|H_0\|=\frac{3}{2} J$ and $\|V\|=2 B$.
The eigenvalues and eigenvectors of the unitary operator $U$ are given by
\begin{eqnarray}
\Lambda(U)=\{e^{i(\frac{J}{2}\pm a)}\},\ \ \ \ \ \ \ |v_{\pm}\rangle=(\frac{2B\mp a}{J},1),\nonumber 
\end{eqnarray}
and thus we can easily obtain with projector operators $\Pi_{\pm}=\frac{|v_{\pm}\rangle\langle v_{\pm}^t|}{\|v_{\pm}\|^2}$.
It follows that, using the identities
\begin{align}
[\sigma_x,\sigma_z\sigma_x]&=-2 \sigma_z, \nonumber \\
[\sigma_z,\sigma_x]&=2 i \sigma_y,\nonumber \\
[\sigma_z,\sigma_z\sigma_x]&=2 \sigma_x,\nonumber 
\end{align}
and after some algebra, we obtain\newline
\begin{eqnarray}
W(t\geq t_0)&=&\text{Tr}(\rho_0[U_t,U_t^\dagger H_0])\\
&=&\text{Tr}(\rho_0\hat{\Delta H(t)}),\nonumber 
\end{eqnarray}
where we defined the operator $\hat{\Delta H(t)}$ as
\begin{eqnarray}
\hat{\Delta H(t)}&=&-\frac{4 B J^2 \sin ^2\left(a (t-t_0)\right)}{a^2}\hat \sigma_z\\
&-&\frac{2 B J \sin
   \left(2 a (t-t_0)\right)}{a}\hat \sigma_y\nonumber \\
  &+&\frac{4 B^2 J \left(1- \cos
   \left(2a (t-t_0)\right)\right)}{a^2}\hat \sigma_x\nonumber 
\end{eqnarray}
Since the system in this protocol has only two involved states, we can write 
\begin{eqnarray}
\rho_0=\epsilon_0 I+\epsilon_1 \sigma_x+\epsilon_2 \sigma_y+\epsilon_3 \sigma_z,
\end{eqnarray}
with the constraints $\text{Tr}(\rho_0)=1$. If we impose the constraint, we must impose $\epsilon_0=\frac{1}{2}$, and write
\begin{eqnarray}
\rho_0= \frac{I}{2}+\epsilon_1 \sigma_x+\epsilon_2 \sigma_y+\epsilon_3 \sigma_z.
\end{eqnarray}
The purity of the state is then given by 
\begin{eqnarray}
\text{Tr}(\rho^2)=\frac{1}{2}+2(\epsilon_1^2+\epsilon_2^2+\epsilon_3^2)=\frac{1}{2}+2\|\vec \epsilon\|^2,\nonumber 
\end{eqnarray}
from which it follows that $\|\epsilon\|^2\leq \frac{1}{4}.$
The work can then be written as
\begin{eqnarray}
W(t)&=&%-\frac{8 B J^2 \sin ^2\left(a (t-t_0)\right)}{a^2}\epsilon_3\\
%&-&\frac{4 B J \sin
%   \left(2 a (t-t_0)\right)}{a} \epsilon_2\nonumber \\
%  &+&\frac{8 B^2 J \left(1- \cos
%   \left(2a (t-t_0)\right)\right)}{a^2}\epsilon_1.
%=
4 B J \Big(\frac{\left(4 B \epsilon _1-2 J \epsilon _3\right) \sin ^2\left( at\right)}{a^2} \nonumber \\
&-&\frac{2\epsilon _2
   \sin \left( at \right) \cos \left( at\right)}{a}\Big).
\end{eqnarray}
It is easy to see that work is zero only if $4 B \epsilon_1-2 J \epsilon_3=0$ and $\epsilon_2=0$.

Let us now focus on the upper bound. In order to calculate the upper bound, we first evaluate
\begin{eqnarray}
& &\mathbb C_{U_t}(\rho_0)=\|[\Pi_+,\rho_0]\|^2+\|[\Pi_-,\rho_0]\|^2\nonumber \\
&&=\frac{4 \left(\epsilon _2^2 \left(4 B^2+J^2\right)+4 B^2 \epsilon _1^2-4 B J \epsilon _3 \epsilon _1+J^2 \epsilon _3^2\right)}{4
   B^2+J^2}.\nonumber 
\end{eqnarray}
It follows that bound A for the work is given by
\begin{eqnarray}
& &|W(t)|\leq  \nonumber \\
& &\sqrt{8 J^2\frac{ \left(\epsilon _2^2 \left(4 B^2+J^2\right)+4 B^2 \epsilon _1^2-4 B J \epsilon _3 \epsilon _1+J^2 \epsilon _3^2\right)}{4
   B^2+J^2}},\nonumber
\end{eqnarray}
from which we see that the upper bound is of the same order of magnitude as the work itself. Also, it is easy to see that if $\epsilon_2=4 B \epsilon_1-2 J \epsilon_3=0$, then $\epsilon _2^2 \left(4 B^2+J^2\right)+4 B^2 \epsilon _1^2-4 B J \epsilon _3 \epsilon _1+J^2 \epsilon _3^2=0$, which implies that the maximum work is also zero, showing that the bounds are tight for this set of parameters.

We can also calculate the power. We observe that
\begin{widetext}
\begin{eqnarray}
P(t)=\frac{dW(t)}{dt}&=&8 B J \Big(\frac{\left(J \epsilon _3-2 B \epsilon _1\right) \sin \left(2 t \sqrt{4 B^2+J^2}\right)}{\sqrt{4 B^2+J^2}} -\epsilon _2
   \cos \left(2 t \sqrt{4 B^2+J^2}\right),\Big)
\end{eqnarray}
\end{widetext}
while our bound A on the power is given by
$\|H_0\|\cdot \|V\|\sqrt{ \text{r}([V,\rho_t]) \cdot \mathbb C_V(\rho_t)}$.
We have $\Pi_V^1=(1,0)^t \otimes (1,0)$ and $\Pi_V^2=(0,1)^t \otimes (0,1)$, or
\begin{eqnarray}
    \Pi^1_V=\begin{pmatrix}
    1 & 0\\
    0 & 0\\
    \end{pmatrix},
    \Pi^2_V=\begin{pmatrix}
    0 & 0\\
    0 & 1\\
    \end{pmatrix},
\end{eqnarray}
It follows that
\begin{eqnarray}
\mathbb{C}_V(\rho_0)=2(\epsilon _1^2+\epsilon _2^2).
\end{eqnarray}
$P_C\leq2* 3/2 J*  2B* \sqrt{2(2 \epsilon _1^2+2\epsilon _2^2}=12 BJ \sqrt{ \epsilon _1^2+\epsilon _2^2)}$, which is also of the same order as the power.

\subsection{The anisotropic XY model}
 
With increasing experimental control over larger unitary quantum systems, it becomes more pressing to study the thermodynamics of many-body quantum systems. 
Theoretical studies have shown advantages by exploiting collective effects in quantum batteries~\cite{battp,ModiPRL2017,polini,
adc,marti2017,ghosh}. 
Here, we apply the bounds on work extraction to a simple many-body spin system described by the XY model~\cite{xymodel2,xymodel3,xymodel}, which can be  investigated experimentally \cite{xymodelexp1,xymodelexp2}.
%XXX refs \cite{campaioli,Modiarxiv2017} adc, check new XXX xxx advantage by collective global operations
%]]]}

The (anisotropic) transverse-field XY model in one dimension is a well-known spin model in Statistical Mechanics. One of its advantages is that properties of the ground
and excited states  are known exactly. The transverse XY Ising model has an interesting phase diagram in which also quantum phase transitions are present.
 In recent years the transverse field XY model has been studied in relation to quantities 
of interest in quantum information theory, such as entanglement and quantum discord. Also, it has been shown that fidelity measures present signatures of QPTs. %Here, we are in particular interested in its features related to decoherence, in which the quantum system coherence changes due to its interaction. 

The Hamiltonian of the anisotropic XY spin chain reads
\begin{equation}\label{eq:XYmodel}
    H = -   \frac{1}{2}\sum_{i=1}^N\left( \frac{1+\eta}{2}\sigma_i^x\sigma_{i+1}^x + \frac{1-\eta}{2}\sigma_i^y\sigma_{i+1}^y +h\sigma_i^z \right),
\end{equation}
where $h$ is the external magnetic field, and $\eta$ is the anisotropy parameter. We also assume perodic boundary condition.

For any value of $\eta, h$,  the XY model can be diagonalized using the Jordan-Wigner transformation:
\begin{equation}
    \sigma_i^z = 1-2c^\dag_ic_i,\ \sigma_i^- = c_i^\dag e^{i\pi\sum_{i=1}^{i-1}c^\dag_j c_j}. 
\end{equation}
The Hamiltonian is then diagonalzed in the form
\begin{equation}
    H =  \sum_{k>0} \Lambda_k\left( \gamma^\dag_k \gamma_k + \gamma^\dag_{-k} \gamma_{-k} -1 \right).
\end{equation}
Here, the fermionic operator $\gamma_k$ is defined by the Bogoliubov transformation of the Fourier transformed operators, $c_k$ i.e.
\begin{equation}
    c_k = \cos{\theta_k}\gamma_k + i \sin{\theta_k}\gamma^\dag_{-k},
\end{equation}
with the dispersion ralation given by $\Lambda_k = \sqrt{\epsilon_k^2 + \eta^2\sin^2{k}}$, $\epsilon_k = h - \cos{k}$, and the angle $\theta_k = \tan^{-1}[(\eta\sin{k})/(\epsilon_k+\Lambda_k)]$.
\subsubsection{Work and power}
We take the XY model Hamiltonain without external field $h$ as the internal Hamiltonian of the battery, and the battery is initially prepared in the ground state of its internal Hamiltonian. 
The charging process is achieved by turning on a constant field, i.e., via a standard quantum quench. In the following, we use superscript $(1)$ to label the operators with initial parameters, and $(2)$ (or no superscript for simplicity) to label the operators with quenched parameters. 
%%%%%%%%%%
\begin{figure}
    \centering
    \includegraphics[width=0.48\textwidth]{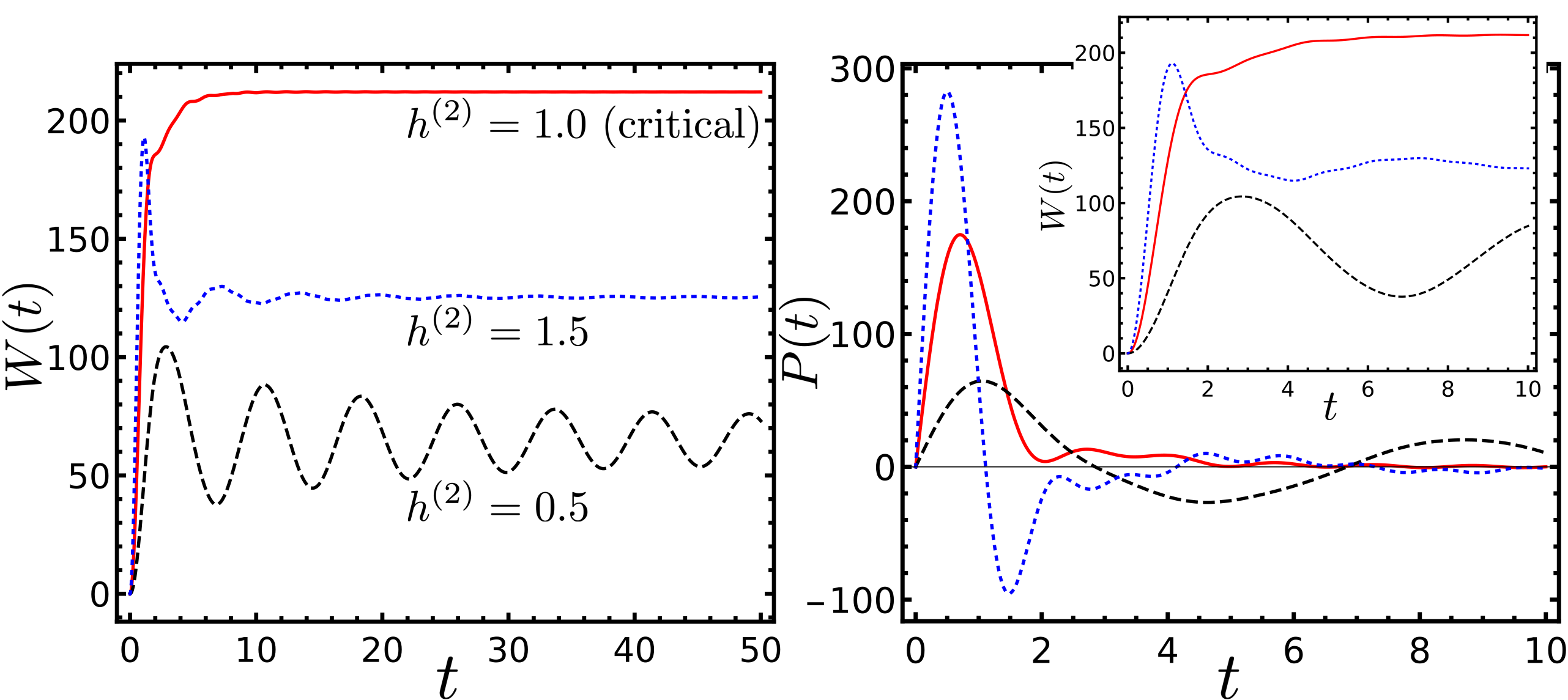}
    \caption{Work $W(t)$ (left) and power $P(t)$ (right) after a quantum quench of the anisotropic XY model (\ref{eq:XYmodel}) for $N=1000$ spins, with the anisotropic parameter $\eta = 0.5$. The battery internal (unquenched) Hamiltonian has zero field, $h^{(1)} = 0$. Colors label various quenching field $h^{(2)}$ at both non-critical (black dashed and blue dotted) and critical values (red solid).}
    \label{fig:figureXY}
\end{figure}
%%%%%%%%%%%%%
The initial and quenched fermionic operators $\gamma_k$ are linked through
\begin{equation}
    \gamma_k^{(1)} = \cos{\chi_k} \gamma_k^{(2)} + i\sin{\chi_k} \gamma_k^{(2)\dag},
\end{equation}
where $\chi_k \equiv \theta_k^{(2)} - \theta_k^{(1)}$. This allows us to express the initial Hamiltonian in terms of the operators with the quenched parameters, i.e.,
\begin{equation}\label{eq:unquenchH}
\begin{aligned}
        H^{(1)} &=  \sum_{k>0} \Lambda_k^{(1)} [ \cos^2{\chi_k}(\gamma_k^\dag\gamma_k+\gamma_{-k}^\dag\gamma_{-k}) + \\ 
       &\ \ \  + \sin^2{\chi_k}(\gamma_k\gamma^\dag_k+\gamma_{-k}\gamma^\dag_{-k}) + \\ &\ \ \  + 2i\sin{\chi_k}\cos{\chi_k} (\gamma_k\gamma_{-k}+\gamma^\dag_k\gamma^\dag_{-k}) - 1].
\end{aligned}
\end{equation}
The initial ground state of the battery internal Hamiltonian can also be written in terms of the quenched operators,
\begin{equation}
    |\psi(0)\rangle = \prod_{k>0} (\cos{\chi_k} - i\sin{\chi_k}\gamma_k^\dag\gamma_{-k}^\dag)|0_k\rangle,
\end{equation}
where $|0_k\rangle$ is the vaccum state of $\gamma_{\pm k}$.
The time evolved state is then,
\begin{equation}\label{eq:quenchwavef}
    |\psi(t)\rangle = \prod_{k>0} (\cos{\chi_k}e^{i\Lambda_kt} - i\sin{\chi_k}e^{-i\Lambda_kt}\gamma_k^\dag\gamma_{-k}^\dag)|0_k\rangle.
\end{equation}

Now, the time-dependent work and power can be computed explicitly, i.e.,
\begin{equation}\label{eq:XYwork}
    W(t) = 2\sum_{k>0} \Lambda_k^{(1)} \sin^2{2\chi_k} \sin^2{\Lambda_k^{(2)}t},
\end{equation}
and
\begin{equation}\label{eq:XYpower}
    P(t) = 2\sum_{k>0} \Lambda_k^{(1)}\Lambda_k^{(2)} \sin^2{2\chi_k}\sin{2\Lambda_k^{(2)}t}.
\end{equation}

For non-zero anisotropic paramaters $\eta$, the XY model exhibits two regions of criticality at $h=\pm 1$. {Figure~\ref{fig:figureXY} depicts the charging work (\ref{eq:XYwork}) and power (\ref{eq:XYpower}) with the battery Hamiltonian quenched to critical and non-critical values. The battery reaches a more stable charging process and a higher work with a critical quench. However, even though non-critical quenches finally charges the battery to a lower work, the initial power is higher for larger quench fields. Therefore, an optimal protocol could use an initial high field to fast charge the battery, and then switch to a critical field to maintain the maximal work.}

\subsubsection{Upper bounds}
We now apply the some of the upper bounds derived in the previous sections to the work and power of the anisotropic XY model. 

The first example we studied is the bound (C) for work,  which involves the coherence of the internal Hamiltonian $H_0 = H^{(1)}$ in the basis of the evolution operator. The coherence of $H_0$ can be computed as well, i.e.,
\begin{equation}
\begin{aligned}
        & C_{U_t}(H_0)  =  \\
        & \sum_\alpha \left[\sum_{k>0}\delta_{\alpha_k,\alpha_{-k}} (-1)^{\beta_k(\alpha)} \Lambda_k^{(1)} (1+\cos{(2\chi_k)}) \right. \\
       &\left. \times \sum_{k>0} \delta_{\alpha_k,\alpha_{-k}} (-1)^{\beta_k(\alpha)} \Lambda_k^{(1)} (1-\cos{(2\chi_k)}) \right],
\end{aligned}
\end{equation}
where the outer summation ranges over all possible configurations of the strings $\alpha = \{\alpha_k,\alpha_{-k}\}_{k>0}$ and $\alpha_{\pm k}$ is either $0$ or $1$. $\delta$ is the Kronecker delta. Also,
\begin{equation}
    \beta_k(\alpha) = \left\{
    \begin{matrix}
    0, \  \text{if}\ \alpha_k = \alpha_{-k} = 1,\\
    1, \  \text{if}\ \alpha_k = \alpha_{-k} = 0.
    \end{matrix}
    \right.
\end{equation}
This result gives a quantitative evaluation of the energy extracted in a quantum many-body quench since the work extracted is basically upper bounded by twice the square root of the coherence of $H_0$ (See Eq.~(\ref{eq:goodbadugly})). As confirmed by our numerical evaluation, this upper bound increases exponentially fast in the number of spins. Consequently, for the quench parameters in Fig.~\ref{fig:figureXY}, this bound is larger than the exact work for a few orders of magnitude. The exponential scaling is more apparent for the upper bounds of the power. Hence, these upper bounds can be very loose for many-body systems. %However, they could be very useful for small systems which are more sensitive to the structure of the system Hamiltonian as detected by the coherence in corresponding the upper bounds.

The more interesting case is the upper bound (A) of the work in terms of the coherence of the initial state in the basis of the time evolution operators. The the quench Hamiltonian is time-independent, hence the eigen-basis of the unitary evolution operator $U$ is the basis of the Hamiltonian itself, i.e.,
\begin{equation}\label{eq:projectbasis}
    \Pi_\alpha = \prod_{k>0} (\gamma_k^\dag)^{\alpha_k}(\gamma_{-k}^\dag)^{\alpha_{-k}}|0\rangle\langle 0|\gamma_k^{\alpha_k}\gamma_{-k}^{\alpha_{-k}}.
\end{equation}
Here, $\alpha_k$ is either $0$ or $1$, and the string $\alpha = \{\alpha_k,\alpha_{-k}\}_{k>0}$ labels posibble configurations of all $\alpha_k$'s. There are $2^N$ possible configurations corresponding to the $2^N$ projectors. 

\begin{figure}[t!]
    \centering
    \includegraphics[width=0.45\textwidth]{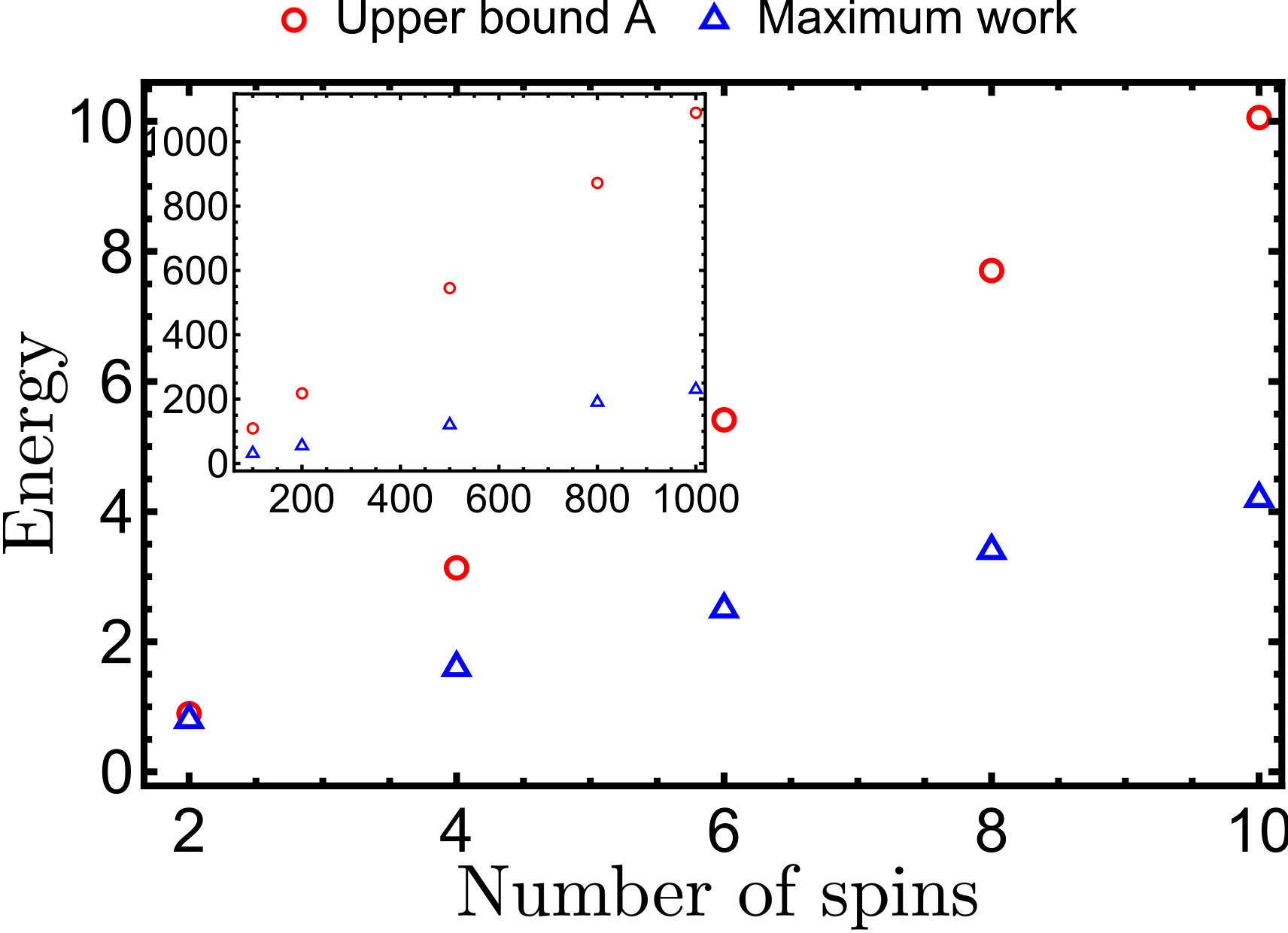}
    \caption{{Comparisons between the upper bound (Eq.~(\ref{eq:workboundA})) of the work for the anisotropic XY model and the maximum value of the exact work solution (\ref{eq:XYwork}).}}
    \label{fig:bounds}
\end{figure}

For each given projector, its average over the initial state $\rho_0 = |\psi(0)\rangle\langle\psi(0)|$ is
\begin{equation}
\begin{aligned}
    &\langle\psi(0)|\Pi_\alpha|\psi(0)\rangle\\
    =&  \prod_{k>0}\left(\cos^2{\chi_k}\delta_{\alpha_k,0}\delta_{\alpha_{-k},0} + \sin^2{\chi_k}\delta_{\alpha_k,1}\delta_{\alpha_{-k},1} \right).
\end{aligned}
\end{equation}
The coherence is then
\begin{equation}
    \begin{aligned}
       C_{U_t}(\rho_0)
       =&\text{Tr}\rho_0^2 - \sum_\alpha|\langle\psi(0)|\Pi_\alpha|\psi(0)\rangle|^2 \\
       =& 1 - \prod_{k>0}\left(\cos^4{\chi_k} + \sin^4{\chi_k} \right) .
    \end{aligned}
\end{equation}
The maximum eigenvalue of initial Hamiltonian is
\begin{equation}
    E^{(1)}_{max} = \sum_{k>0} \Lambda_k^{(1)}. 
\end{equation}
Together with the coherence of the initial state, we get the upper bound (A) of the work,
\begin{eqnarray}\label{eq:workboundA}
    W(t) 
    &\le&2\sqrt{2}|\sum_{k>0} \Lambda_k^{(1)}| \sqrt{1 - \prod_{l>0}\left(\cos^4{\chi_l} + \sin^4{\chi_l} \right)}\nonumber \\
\end{eqnarray}
Since $\chi_k \equiv \theta_k^{(2)} - \theta_k^{(1)}$ is cannot be zero for all $k's$, unless the quenched Hamiltonian is the same as the initial Hamiltonian, the product term in the above bound vanishes exponentially fast with the system size. Hence, this upper bound of the work is only determined by the largest eigenvalue of the internal Hamiltonian, i.e., its operator norm, which scales linearly in the number of spins, and again can be very loose for many-body systems.  However, they could be very useful for small systems which are more sensitive to the coherence of the initial state. In this case the product term in the above equation can have non-trivial contributions. {Fig.~\ref{fig:bounds} shows the upper bound (A) and the maximum work extracted from the exact solution (\ref{eq:XYwork}) at various number of spins. The bound becomes tighter for smaller particle numbers.}

%%%%%%%%%%

\subsection{Spin-Boson model}
As an application of the bounds in the open system, we consider the simplest exactly solved model, the spin-boson model \cite{Vernon,Caldeira,Palma}. The spin-model model describes the precession of  the two-level system of a spin  in an open environment. The open environment is described by a reservoir of Harmonic oscillators. The model is well known because it serves the purpose of describing decoherence in a simple exactly solvable setting \cite{schlosshauer}.

The total Hamiltonian of the system  is given by
\begin{eqnarray}
    H&=&\underbrace{\frac{1}{2}( \omega_0\sigma_z-\Delta_0 \sigma_x)}_{H_0} +\sum_k \omega_k b_k^\dagger b_k \nonumber \\
    & &+\sum_k \sigma_z (g_k b_k^\dagger+g_k^* b_k).
\end{eqnarray}
The constants $\omega_0$ and $\omega_k$ are the level spacing of the two-level system and the frequencies of the bosonic degrees of freedom respectively, with $[b_k,b_{k^\prime}^\dagger]=\delta_{k,k^\prime}$. The parameter $\Delta_0$ is associated to the tunneling between the two states.

Here, we consider $\mathcal S$ to be the two-level system. Under the condition of markovianity, the master equation for the spin-boson system can be written in the form \cite{schlosshauer}
\begin{eqnarray}
    \frac{d}{dt} \rho_{\mathcal S}&=&\frac{i}{\hbar} [\rho_{\mathcal S},H_0]+\gamma \sigma_z \rho_{\mathcal S} \sigma_z^\dagger-\frac{\gamma}{2} \sigma_z^\dagger \sigma_z \rho_{\mathcal S}-\frac{\gamma}{2} \sigma_z^\dagger \sigma_z\nonumber \\
    &=&\frac{i \omega_0}{2\hbar} [\rho_{\mathcal S},\sigma_z]-\frac{i\Delta_0}{2\hbar}[\rho_{\mathcal S},\sigma_z]+\gamma \sigma_z \rho_{\mathcal S} \sigma_z-\gamma \rho_{\mathcal S},\nonumber \\
\end{eqnarray}
where we used $\sigma_z^2=I$.
where the single Lindblad operator are given by $L_1=L_1^\dagger=\sigma_z$, and thus is Hermitian. In the derivation of the reduced equation above the assumption of a thermal reservoir is made, with the constant $\gamma>0$ containing the spectrum of the bath.

For the reduced system, it is easy to see that the energy storage is given by
\begin{eqnarray}
    E(t)=\frac{\omega_0}{2}\text{tr}(\rho_{\mathcal S} \sigma_z)
-\frac{\Delta_0}{2}\text{tr}(\rho_{\mathcal S} \sigma_x).\end{eqnarray}
If we define
\begin{eqnarray}
    \rho_{\mathcal S}=\begin{pmatrix}
    \rho_{11} & \rho_{12} \\
    \rho_{21} & \rho_{22}
    \end{pmatrix} \rightarrow E(t)&=&\frac{\omega_0}{2} (\rho_{11}-\rho_{22})\nonumber \\
    &-&\frac{\Delta_0}{2} (\rho_{12}+\rho_{21}).\nonumber \\
\end{eqnarray}

The evolution of the system is given by the following general dynamical equations:
\begin{eqnarray}
    \dot \rho_{11}&=&-\frac{1}{2} i \Delta_0  \left(\rho _{12}-\rho _{21}\right),\nonumber \\
    \dot \rho_{22}&=&-\frac{1}{2} i \Delta_0  \left(\rho _{21}-\rho _{12}\right),\nonumber\\
    \dot \rho_{12}&=&-\gamma  \rho _{12}-\frac{1}{2} i \Delta_0  \left(\rho _{11}-\rho
   _{22}\right)-\frac{1}{2} i \rho _{12} \omega_0,\nonumber \\
    \dot \rho_{21}&=&-\gamma  \rho _{21}-\frac{1}{2} i \Delta_0  \left(\rho _{22}-\rho
   _{11}\right)+\frac{1}{2} i \rho _{21} \omega_0,\nonumber \\
\end{eqnarray}
from which it is immediate to see that the trace of the density matrix is preserved, but the off-diagonal terms change due to decoherence arising from the interaction with the bath.
Substituting the equations above into $\frac{dE}{dt}$ we obtain
\begin{eqnarray}
    \frac{dE}{dt}&=&-\frac{\Delta_0}{2} \Big(\rho _{12} \left(-\gamma -\frac{i \omega_0 }{2}\right)+\rho _{21} \left(-\gamma
   +\frac{i \omega_0 }{2}\right)\Big)\nonumber\\
   &=&\Delta_0 \text{Re}\Big(\rho _{12} \left(\gamma +\frac{i \omega_0 }{2}\right)\Big) \nonumber \\
   &=&\Delta_0 \gamma \text{Re}(\rho_{12})-\frac{\Delta_0 \omega_0}{2} \text{Im}(\rho_{12}),
\end{eqnarray}
from which we see instead that a change in energy is associated to the tunneling, and is zero otherwise. It is also easy to see that the energy storage changes when coherence is present, due to the proportionality of its derivative with $\rho_{12}$.

We now wish to apply the bounds of Sec.~\ref{sec:lindblad}, in particular Eq.~(\ref{eq:opensysbound}), in which only term $W_B$ contributes:
\begin{eqnarray}
    \frac{|\Delta E(t)|}{\|H_0\|}&\leq &  3\gamma  t  \text{sup}_{\tau\in [0,t]} \sqrt{r_n \mathbb C_{L_n}\big(\rho_{\mathcal S}(\tau)\big) }.  \nonumber 
\end{eqnarray}
To this end, we first calculate various quantities that are related to the bound.
The eigenvalues of $\Lambda(H_0)=\{\pm \frac{1}{2} \sqrt{\Delta_0 ^2+\omega_0 ^2}\}$. Thus we have $\|H_0\|=\frac{1}{2} \sqrt{\Delta_0 ^2+\omega_0 ^2}\approx \Delta_0/2$, while $\|L_1^2\|=1$ and $l_1=1$. Also, both $R_n\leq 4$ and $r_n\leq 2$. The spectral decomposition of $\sigma_z$ is given by 
\begin{eqnarray}
    \Pi_1=\begin{pmatrix}
    1 & 0\\
    0 & 0\\
    \end{pmatrix},
    \Pi_2=\begin{pmatrix}
    0 & 0\\
    0 & 1\\
    \end{pmatrix},
\end{eqnarray}
from which, after an immediate calculation, we get
\begin{eqnarray}
    \sqrt{\mathbb C_{L_1}(\rho_{\mathcal S})}= \sqrt{2 (|\rho_{12}|^2+|\rho_{21}|^2)}.
\end{eqnarray}
Using the fact that $\rho_{21}=\rho_{12}^*$, 
assuming $\hbar=1$, and applying the upper bound of Eq.~(\ref{eq:opensysbound}), we obtain
\begin{eqnarray}
    \frac{|E(t)-E_0|}{t}&\leq& \Delta_0 3\sqrt{2} \gamma sup_{\tau\in [0,t]} |\rho_{12}(\tau)|\nonumber \\
    &\leq& 4.24(2) \Delta_0 \gamma |\rho_{12}^0|.
\end{eqnarray}
Let us now make some plots in the regime of strong decoherence and tunneling, in which we have $\gamma\gg \Delta_0\gg \omega_0$.
In this case we have $\rho_{12}(t)=\rho_{21}^*(t)\approx \rho_{12}^0 e^{-(i \frac{\omega_0}{2}+\gamma) t}$, and thus,
\begin{eqnarray}
    E(t)-E_0\approx  |\rho_{12}^0|\Delta_0\frac{2 e^{-\gamma  t} \left(-  \cos \left(\frac{t \omega_0
   }{2}\right)+  e^{\gamma  t}\right)}{2  }.\nonumber \\
\end{eqnarray}
At the first order in $t$, we have instead the exact result
\begin{eqnarray}
   \frac{| E(t)-E_0|}{t}\approx  0.25  \Delta_0\gamma |\rho_{12}^0|   +O(t)\nonumber.
\end{eqnarray}
We thus see that for short times the bound has the same functional form as the actual first order approximation of the energy, and the two constants are both of order one.

\section{Conclusions}

In this paper, we studied and have made an explicit  connection between work and coherence in a quantum battery. 
In particular, we have provided a quantitative framework for evaluating upper bounds to energy storage in quantum batteries, based on the notion of Hilbert-Schmidt density matrix coherence, and its generalization to other operators like the Hamiltonian used to measure the energy of the system, and the internal Hamiltonian $H_0$. 
We derived two key results. The first one is that the charging (or discharging) of a quantum battery is  governed by the amount of non commutativity of both the density matrix of the system and the internal Hamiltonian in the basis given by the spectral decomposition of the interaction.  We provided bounds which clarify the intricate interplay between coherence of the density matrix, of the Hamiltonian and the interaction in order for work to be performed by a quantum battery. The amount of non commutativity is quantified by using the Frobenius norm, namely by summing the modulus squares of all off-diagonal elements of an operator in the desired basis. For a state, this is a measure of coherence, and is bounded by one. For a generic operator, e.g. $H_0$, this form of coherence can be very large and 
scale with the size of the Hilbert space in certain cases, rather than the number of particles in the system. As such, some of our bounds, if taken as they are, can be very loose {for large systems. 
Nonetheless, this also implies that some of our bounds can capture} the behaviour of small systems well. We have tested these bounds on two exactly solvable closed quantum models, i.e. a 4-level system and the anisotropic XY model. In the former, our bounds are fairly tight, while in the latter we show under which conditions the operator coherence is small enough in order to obtain a tight bound. 

The second result is the extension to the case of generic quantum channels, e.g., open quantum systems, both in terms of Kraus and Lindblad operators.
 In order to see how these bounds apply quantitively, we have studied ensembles of quantum batteries and the Spin Boson model, showing the role of coherence in charging such battery model. 
In the case of open systems we have focused on Hermitian Kraus and Linblad operators, but some of these bounds can be extended to non-Hermitian ones, with some technical challenges. This will be the subject of future investigations. In perspective, we find that the results of this paper open a certain number of interesting questions. Since in an open quantum system coherence is typically exponentially suppressed, we are interested in showing how decoherence free subspaces \cite{zanardidfs, violadfs, lidardfs} can be used to obtain more efficient quantum batteries. In the spirit of the typicality arguments used both in \cite{Caravelli1, zanardicoh,isospec2} we can ask how typical quantum maps can be used to exchange energy. Finally, an important generalization would be to take in consideration the entropy change in open quantum systems and extend these results to the free energy available to a quantum battery.

{In the case of many-body quantum batteries, it is natural to connect coherence to entanglement within the constituents of the battery.
Connections between the charging of quantum batteries and their entanglement have been studied in~\cite{alicki,Modiarxiv2017,ModiPRL2017,battp,LewensteinBatteries18}. 
%In~\cite{LewensteinBatteries18}, it was shown that that there is a relationship between power and entanglement. %In particular,  a Hamiltonian which can is k-qubit entangled, then the instantaneous power is upper bounded by the Fisher information in energy eigenspace. 
In this paper we have focused on an Frobenius measure of coherence. While it does not satisfy all the coherence monotones axioms defined in \cite{plenio}, 
such coherence measure is connected to entanglement, in the sense that the more a state is entangled the less coherence can be stored in certain local parts of a system, which can be seen analyzing the Frobenius measure of coherence, via a Schmidt decomposition~\cite{hammaloccoh}. Thus, the higher the entanglement in a certain system the more coherent the state can be. In addition, some of our bounds could possibly be extended to $l_1$ measures of coherence, which will be to focus of future investigations. If this is case, it is known that from the point of view of resource theory that entanglement and coherence are quantitatively equivalent \cite{streltsov2}. Thus, while our bounds focused on the relation between energy storage and coherence, a complementary picture can be obtained in terms of entanglement.}

{
Moreover, in \cite{Caravelli1, isospec1} it has been shown that, in the context of random quantum batteries, there is a quantum advantage with respect to classical devices due to the behavior of the spectral gaps in the eigenvalues of the evolution operator $U$. These gaps are relevant when the initial state populates both eigenstates of $U$, and in turn this contributes to the coherence of $U$ in the basis of the initial state. It would  be interesting to see whether one can bound the quantum advantage of a battery in terms of coherence, which would result in a guide to designing superior devices at the microscopic level. }

\section*{Acknowledgments} We acknowledge the support of NNSA for the U.S. DoE at LANL under Contract No. DE-AC52-06NA25396. A.H. acknowledges  support from
NSF award number 2014000. FC was also financed via DOE-LDRD grants PRD20170660 and PRD20190195.
L.P.G.P. acknowledges partial support by AFOSR MURI project ``Scalable Certification of Quantum Computing Devices and Networks'', 
DoE ``Fundamental Algorithmic Research for Quantum Computing (FAR-QC)",
DoE ASCR FAR-QC (award No. DE-SC0020312), DoE BES Materials and Chemical Sciences Research for Quantum Information Science program (award No. DE-SC0019449), DoE ASCR Quantum Testbed Pathfinder program (award No. DE-SC0019040), NSF PFCQC program, AFOSR, ARO MURI, AFOSR MURI, and NSF PFC at JQI.
B.Y. also acknowledges support from the U.S. DoE, Office of Science, Basic Energy Sciences, Materials Sciences and Engineering Division, Condensed Matter Theory Program, and partial support from the Center for Nonlinear Studies.

\bibliographystyle{plainnat}
\bibliography{references}

\onecolumn\newpage
\appendix

\section{Work and von Neumann's trace inequality}\label{sec:appvn}
Let us now consider another set of bound for the work, based on von Neumann's trace inequalities. Consider two operators $\mathcal A$ and $\mathcal B$. Let $\alpha_i$ and $\beta_i$ be the ordered set of singular values of $\mathcal A$ and $\mathcal B$, e.g. $\alpha_{i-1}\geq \alpha_{i}$, and $\beta_{i-1}\geq\beta_{i}$.
Then we have
\begin{eqnarray}
    |\tr{\mathcal A\mathcal B}|\leq \sum_{i=1}^n \alpha_i \beta_i,
\end{eqnarray}
and if $\mathcal A$ and $\mathcal B$ are Hermitian, then 
\begin{eqnarray}
    \sum_{i=1}^n \alpha_i \beta_{n-i+1}\leq \tr{\mathcal A\mathcal B}\leq \sum_{i=1}^n \alpha_i \beta_i.
\end{eqnarray}
It follows that, if we use the equalities from the previous section:
\begin{equation}
\label{eq:work}
    W=\underbrace{\text{Tr}(U_t^\dagger H_0 [\rho_0,U_t])}_{A}=\underbrace{\text{Tr}(\rho [U_t,U_t^\dagger H_0)]}_{B}=\underbrace{\text{Tr}(U_t[U_t^\dagger H_0,\rho_0])}_{C},
\end{equation}
we have three pairs of operators: $\mathcal A_A=U_t^\dagger H_0$, $\mathcal B_A=[\rho,U_t]$; $\mathcal A_B=\rho$, $\mathcal B_B=[U_t,U_t^\dagger H_0]$ and $\mathcal A_C=U_t$, $\mathcal B_C=[U_t^\dagger H_0,\rho]$.

Let us thus consider the singular values of the three pairs. Given an operator $Q$, the singular values square are the (square roots of the) eigenvalues of the operator $Q_2=Q^\dagger Q$. 
Von Neumann's trace inequality can be applied
to equality $C$ as both $\rho$ and $[U_t,U_t^\dagger H_0]=H_0-U_t^\dagger H_0 U_t$ are Hermitian operators.
The singular values of  $\rho$ are the eigenvalues of $\rho^2$. Let $d_i$ be the singular values of $\rho$, and $\sigma_i(t)$ be the singular values of $H_0-U_t^\dagger H_0 U_t$.
Then, it follows that
\begin{eqnarray}
   \sum_{i=1}^n d_{n-i+1} \sigma_i(t) \leq |W(t)|\leq \sum_{i=1}^n d_i \sigma_i(t)
\end{eqnarray}

We can perform another upper bound:
$W(t) < \sum_i d_i \sigma_i(A+B)$ where A is $H_0$ and $B=-U^\dagger H_0 U$
now, we can use Weyl's inequality for the singular values of the sum of Hermitian matrices. This is
$$W(t) < \text{max}\ d  \sum_i \sigma_i(A+B)<\text{max}\ d  \sum_i( \sigma_i(A)+\sigma_i(U^\dagger A U))< 2 \text{max}\ d  \sum_i \sigma_i(A)< 2 n d\ \text{max}_i\ \sigma_i(H_0)$$

%%%%

One interesting comment is that both $\rho$ and $[U_t,U_t^\dagger H_0]$ are Hermitian operators. Thus, following von Neumann's trace inequality, we can both upper and lower bound the work with
\begin{eqnarray}
   \underline W=\sum_{i=1}^n d_{n-i+1} \sigma_i(t) \leq |W(t)|\leq \sum_{i=1}^n d_i\sigma_i(t)=\overline W 
   \label{eq:c2}
\end{eqnarray}
where $\sigma_i(t)$ are the singular values of $\hat \epsilon=H_0-U_t H_0 U_t^\dagger$, while $d_i$ are the singular values of $\rho$. It can be immediately seen that the inequality of Eq.~(\ref{eq:c2}) is tighter.

These bounds can be further simplified as follows. Consider the upper 
bound $\overline W=\sum_{i=1}^n d_i\sigma_i(t)$. We can further upper bound it via
\begin{eqnarray}
    \bar W\leq \bar d \sum_j \sigma_j(t),
\end{eqnarray}
where $\bar d=\text{max}_i d_i$. We note that numerically the upper bound above is hard to calculate because of necessity to diagonalize 3 matrices.
Now, for the singular values of a sum of Hermitian matrices we have $\sum_j \sigma_j(A+B)\leq \sum_j (\sigma_j(A)+\sigma_j(B))$ \cite{HJ91}. Now note that $B=-U^\dagger A U$ where $U$ is a unitary transformation, and thus $\sigma_i(U^\dagger A U)=\sigma_i(A)$. It follows that 
\begin{eqnarray}
    \bar W\leq 2\ n\ \bar d\ \bar \sigma(H_0)
\end{eqnarray}
where $n$ is the dimension of the Hilbert space and $\bar \sigma(H_0)=\text{max}_i \sigma_i(H_0)$.
We thus obtain the upper bound
\begin{eqnarray}
    |W(t)|\leq 2\ n\ \bar d\ \bar \sigma(H_0).
\end{eqnarray}
which only requires to solve two maximum eigenvalues problems.

\section{Norms definitions and upper bounds}
\label{sec:appdef}
In the paper we use various norms, so for the sake of clarity we define the following quantities.
Given a matrix $A$, we define the \textit{2-norm}  as $\|A\|_2=\sup_i \sigma_i(A)$, where $\sigma_i(A)$ is the i-th singular value, while $\| A\|$. For a square matrix $A$ of size $n$, We then call the Frobenius norm (or Hilbert-Schmidt) norm the following:
\begin{eqnarray}
    \|A\|_F=\sqrt{ \sum_{ij} |a_{ij}|^2}=\sqrt{\tr{A^\dagger A}}=\sqrt{\sum_{i=1}^n \sigma_i^2(A)}.
\end{eqnarray}
A bound for singular value of two matrices will be the following. Consider two matrices $A$ and $B$. %Then, $\sigma_{i+j-1}(A+B)\leq \sigma_i(A)+\sigma_k(B)$.
%It follows that
%\begin{eqnarray} %k=i+j-1
%\sum_{k=1}^n \sigma_k^2(A+B)\leq\sum_{k} %(\sigma_k(A)+\sigma_{k-1}(B))^2\leq
%\end{eqnarray}
First, let us prove a property of the Frobenius norm that will turn useful in the following. If $\text{tr}(A^\dagger B)=\text{tr}(B^\dagger A)=0$, then $\|A+B\|_F^2=\|A\|_F^2+\|B\|_F^2$. In order to see this, notice that $\|A+B\|_F^2=\text{tr}(A+B)^\dagger(A+B)=\|A\|_F^2+\|B\|_F^2+\text{tr}(A^\dagger B+B^\dagger A)$ from which the statement follows.

In general, we have the inequalities
\begin{equation}
 \|A\|_2  \leq  \|A\|_{F}\leq \sqrt{n} \|A\|_2.
\end{equation}
Let us now prove the following statement. In general, one has that $|\text{tr}(AB)|\leq \sqrt{\|A\|_F^2 \|B\|_F^2}$. In this paper we do use  the notation $\|A\|_F$ for the Frobenius norm to avoid confusion with the spectral norm $\|A\|$. A tighter series of inequalities can be however:
\begin{eqnarray}
    |\text{tr}(A^\dagger B)|\leq \sigma_1(A)\sum_j \sigma_j(B)\leq \sigma_1(A)\sum_j \sigma_j(B)\leq \sqrt{n} \sigma_1(A) \sqrt{\sum_j \sigma_j^2(B)}. 
\end{eqnarray}
%Since $\sum_j \sigma_j^2(B)=\|B\|_F^2 $, the inequality above can be tighter by a factor $\sqrt{n}$ if used properly.

The inequality follows from the following two statements, which can be proved using the singular value decomposition. For any matrix $M=A^\dagger B$, we have $|\text{tr}(M)|\leq \sum_i \sigma_i(M)$. Also, $\sigma_i(A^\dagger B)\leq \sigma_1(A) \sigma_i(B)$, which can be proved using the Fischer min-max theorem.

Another inequality in terms of the Frobenius norm for the trace can be derived as follows.
Consider again $|\text{tr}(A^\dagger B)|$. We can write the following H\"{o}lder inequality
\begin{eqnarray}
|\text{tr}(A^\dagger B)|\leq  |\text{tr}(|A^\dagger|^p)|^{\frac{1}{p}}|\text{tr}(|B|^q)|^{\frac{1}{q}} 
\end{eqnarray}
with $1/p+1/q=1$. Then, if we write $p\rightarrow \infty$ (or q), we can write

\begin{eqnarray}
|\text{tr}(A^\dagger B)|\leq \|A\| \|B\|_1
\end{eqnarray}
where $\|A\|$ is the operator norm, and $\|B\|_1$ the 1-norm. We also make use of the  inequality 
\begin{eqnarray}
\|B\|_F \leq \|B\|_1\leq \sqrt{\text{r}(B)} \|B\|_F
\end{eqnarray}
where $\text{r}(B)$ is the rank of $B$. 
The right inequality holds when the singular values of $B$ are all the same, while the left inequality when only one singular value is nonzero.
We can say something more if at least one of the two matrices can be diagonalized, which is our case in the paper.

If $B=U_t^\dagger [\rho_0,U_t]=\rho_t-\rho_0$, then because of the subadditivity of the rank and the fact that a unitary transformation does not change the rank, we have $\text{r}(\rho_t-\rho_0)\leq 2 \text{r}(\rho_0)$. If the state $\rho_0$ is thus pure, we have that $\text{r}(B)=2$, which is of order $1$.

A bound we will also use is the one for the square root. In fact we have that
\begin{eqnarray}
\sum_{i} \sqrt{\lambda_i a_i}\leq\sqrt{\sum_i \lambda_i} \sqrt{\sum_{i} a_i},
\end{eqnarray}
which is due to the concavity of the square root.

\section{Relation between $\mathbb C(\cdot)$ and $\mathcal D(\cdot)$}
\label{sec:appboundsec}
Let us define the overlap in a certain basis defined by projector operators $\Pi_k$ as
\begin{equation}
\label{eq:coherence}
    \mathbb C(X)=\frac{1}{2}\sum_{j=1}^n\|[X,\Pi_j]\|_F ^2.
\end{equation}
If the operator $X$ is the density matrix, $\mathbb C(\rho)$ is exactly the coherence of $\rho$ in the basis given by $\Pi$'s.

First, we show that the function $\mathbb C(X)$ is the coherence. We define the super operator $\mathcal D(\cdot)=\sum_i \Pi_i \cdot \Pi_i$, and the coherence as 
\begin{eqnarray}
    \|X-\mathcal D(X)\|_F^2 &=&\|X-\sum_{i} \Pi_i X \Pi_i\|_F^2  \nonumber \\
    &=&\text{Tr}\Big((X^\dagger- \sum_j \Pi_j X^\dagger \Pi_j)(X-\sum_i \Pi_i X\Pi_i) \Big) \nonumber \\
    &=&\text{Tr}\Big(X^\dagger X-\sum_j \Pi_j X^\dagger \Pi_j X-X\sum_i \Pi_i X\Pi_i + \sum_{ij} \Pi_j X^\dagger \Pi_j \Pi_i X \Pi_i\Big)\nonumber \\
    &=&\text{Tr}\Big(X^\dagger X-\sum_i X^\dagger \Pi_i X \Pi_i \Big).
\end{eqnarray}
Now we have
\begin{eqnarray}
    \mathbb C(X)\equiv \mathbb C_{\Pi}(X)&=&\frac{1}{2} \sum_i \|[X,\Pi_i]\|_F^2 ={\frac{1}{2}}\sum_i \text{Tr}\Big((X\Pi_i-\Pi_i X)^\dagger(X\Pi_i-\Pi_i X) \Big) \nonumber \\
    &=&\frac{1}{2}\sum_{i}\text{Tr}\Big(\Pi_i X^\dagger X \Pi_i- \Pi_i X^\dagger \Pi_i X-X^\dagger \Pi_i X \Pi_i+X^\dagger \Pi_i \Pi_i X\Big)\nonumber \\
    &=&{\frac{2}{2}} \text{Tr}\Big(X^\dagger X-\sum_i X^\dagger \Pi_i X\Pi_i \Big)= \|X-\mathcal D(X)\|_F ^2.
\end{eqnarray}
{It follows that if $X=\rho$, $\mathbb C(\rho)$} is the coherence operator in the basis of $\Pi$'s. Also, we note that, notationally if we write $X_{ij}$ in the basis of $\Pi_i$'s, in general $\sum_{i\neq j} |x_{ij}|^2= \mathbb C_{\Pi}(X)$. If a certain operator $A$  has a spectral decomposition in terms of the projectors $\Pi$'s, we will write with an abuse of notation $\mathbb C_{\Pi}(\cdot)=\mathbb C_{A}(\cdot)$.

Note that $\|[X,\Pi_i]\|_F^2={\cancel{2}(\sum_k |x_{ik}|^2-|x_{ii}|^2)}$. Also, note that
\begin{eqnarray}
\text{Tr}\Big(X^\dagger X-\sum_i X^\dagger \Pi_i X\Pi_i \Big)&=&\text{Tr}\Big(X^\dagger X\sum_i \Pi_i-\sum_i X^\dagger \Pi_i X\Pi_i \Big) \nonumber \\
&=&\sum_i \text{Tr}\Big(X^\dagger X\Pi_i- X^\dagger \Pi_i X\Pi_i \Big)=\sum_i \text{Tr}(X^\dagger[X \Pi_i,\Pi_i])
\label{eq:cohtr}
\end{eqnarray}
which is an expression we will use later.

\section{Coherence bounds}\label{sec:appbounds}
In this section we prove various auxiliary propositions which enter into the coherence bounds lemmas proved below and reported in the main text.

First, we prove the following:
%{Below is correct, double checked}

\textbf{Proposition 1}. 
Assuming that $\Pi_i \Pi_j=\delta_{ij} \Pi_i$,
we have
$$\text{tr}([\Pi_i,A\Pi_i]^\dagger[\Pi_j,A\Pi_j])=\delta_{ij} \|[\Pi_j,A\Pi_j]\|_F^2$$.

\textit{Proof.} Let us now prove some auxiliary properties related to the spectral decomposition. Note that $\Pi_i^2=I$, and $\Pi_i \Pi_j=0$ if $i\neq j$. Then we have
\begin{eqnarray}
\text{tr}([\Pi_i,A\Pi_i]^\dagger [\Pi_j,A\Pi_j])=0
\end{eqnarray}
for any operator $A$. In fact
\begin{eqnarray}
\text{tr}([\Pi_i,A\Pi_i]^\dagger[\Pi_j,A\Pi_j])&=&\text{tr}\Big((\Pi_i A \Pi_i -A \Pi_i)^\dagger(\Pi_j A\Pi_j -A \Pi_j)\Big) \nonumber \\
&=&-\text{tr}\Big((\Pi_i A\Pi_i - \Pi_i A)(\Pi_j A\Pi_j -A \Pi_j)\Big)\nonumber \\
&=&-\text{tr}\Big(\Pi_i A \Pi_i \Pi_j A\Pi_j- \Pi_i A \Pi_i A\Pi_j- \Pi_i A  \Pi_j A \Pi_j +\Pi_i A \Pi_i \Pi_j A\Pi_j\Big)\nonumber\\
&=&-\text{tr}\Big(\Pi_i A \Pi_i \Pi_j A\Pi_j- \Pi_j\Pi_i A \Pi_i A- \Pi_j\Pi_i A  \Pi_j A  +\Pi_j\Pi_i A \Pi_i \Pi_j A\Big).
\end{eqnarray}
Since in every term there is a product of the form $\Pi_i \Pi_j$, the trace is given by
\begin{eqnarray}
\text{tr}([\Pi_i,A\Pi_i]^\dagger[\Pi_j,A\Pi_j])=\delta_{ij} \|[\Pi_j,A\Pi_j]\|_F^2,
\end{eqnarray}
which completes the proof. $\square$

%{Below is correct, double checked}

\textbf{Proposition 2} 
If $\Pi_i\Pi_j=\Pi_{i} \delta_{ij}$, then
\begin{eqnarray}
\|[\Pi_i,B\Pi_j]\|_F^2=\|[\Pi_i,\Pi_j B]\|_F^2
\end{eqnarray}

\textit{Proof}.
Let us evaluate both sides of the equality. On the right hand side we have
\begin{eqnarray}
\text{tr}([\Pi_i,\Pi_j B]^\dagger[\Pi_i,\Pi_j B])&=& \text{tr}\Big((\Pi_i \Pi_j B- \Pi_j B \Pi_i)^\dagger (\Pi_i \Pi_j B- \Pi_j B \Pi_i)\Big) \nonumber \\
&=&\text{tr}\Big((B\Pi_i \Pi_j-\Pi_i B \Pi_j)(\Pi_i \Pi_j B-\Pi_j B \Pi_i)\Big)\nonumber \\
&=&\text{tr}(B^2 \Pi_i \Pi_j \Pi_i \Pi_j-B \Pi_i \Pi_j \Pi_j B \Pi_i-\Pi_i B \Pi_j \Pi_i \Pi_j B+\Pi_i B \Pi_j \Pi_j B \Pi_i) \nonumber \\
&=&\text{tr}\Big(\delta_{ij}B^2 \Pi_i-2\delta_{ij} B \Pi_i B \Pi_i+\Pi_i B \Pi_j B\Big).
\end{eqnarray}
On the left hand side instead we have
\begin{eqnarray}
\text{tr}([\Pi_i,B \Pi_j]^\dagger[\Pi_i,B \Pi_j])&=&\text{tr}\Big((\Pi_i B \Pi_j-B \Pi_j \Pi_i)^\dagger(\Pi_i B \Pi_j-B \Pi_j \Pi_i)\Big)\nonumber \\
&=&\text{tr}\Big((\Pi_j B \Pi_i-\Pi_i \Pi_j B)(\Pi_i B \Pi_j-B \Pi_j \Pi_i)\Big)\nonumber \\
&=&\text{tr}\Big(\Pi_j B \Pi_i \Pi_i B \Pi_j-\Pi_j B \Pi_i B \Pi_j \Pi_i-\Pi_i \Pi_j B \Pi_i B \Pi_j+\Pi_i \Pi_j B^2 \Pi_j \Pi_i\Big)\nonumber\\
&=&\text{tr}\Big(B \Pi_i B \Pi_j-2 \delta_{ij} B \Pi_i B \Pi_j+\delta_{ij} B^2 \Pi_i \Big)
\end{eqnarray}
from which we see that the equality applies. $\square$

%{Below is correct, double checked}
\textbf{Proposition 3} Let $\Pi_i \Pi_j=\delta_{ij} \Pi_i$. Then, if $A$ is Hermitian, we have
\begin{eqnarray}
\|[\Pi_i,A \Pi_i]\|_F^2=\frac{1}{2} \|[\Pi_i,A]\|_F^2.
\end{eqnarray}

\textit{Proof}. Note that, for $A$ Hermitian
\begin{eqnarray}
\text{tr} [\Pi_i,A \Pi_i]^\dagger [\Pi_i,A \Pi_i]
&=&\text{tr}(\Pi_i A \Pi_i -\Pi_i A)(\Pi_i A \Pi_i - A \Pi_i) \nonumber \\
&=&\text{tr}(\Pi_i A \Pi_i A \Pi_i-\Pi_i A \Pi_i A \Pi_i -\Pi_i A \Pi_i A \Pi_i+\Pi_i A^2 \Pi_i)\nonumber \\
&=&\text{tr}(A^2\Pi_i-\Pi_i A \Pi_i A).
\end{eqnarray}
A rapid calculation shows that instead 
\begin{eqnarray}
\text{tr} [\Pi_i,A]^\dagger [\Pi_i,A]&=&-\text{tr}\Big(\Pi_i A \Pi_i A- A^2 \Pi_i-\Pi_i A^2+A \Pi_i A \Pi_i \Big)=2 \text{tr}(A^2\Pi_i-\Pi_i A \Pi_i A),
\end{eqnarray}
which completes the proof. $\square$

From the Proposition above, it follows that $\sum_{i} \|[\Pi_i, A \Pi_i]\|_F^2=\mathbb C_{\Pi}(A) $.

%{below is correct, double checked}
\textbf{Proposition 4} Let $\Pi_i\Pi_j=\delta_{ij} \Pi_i$ and $B$ Hermitian. Then
\begin{eqnarray}
\text{tr}([\Pi_i,\Pi_j B]^\dagger[\Pi_a, \Pi_b B])=\delta_{ja}\delta_{ab}\text{tr}\Big( B [B\Pi_i, \Pi_a]\Big)+(\delta_{ia}\delta_{jb} -\delta_{ji} \delta_{jb})\text{tr}( B \Pi_j B \Pi_a)
\end{eqnarray}
\textit{Proof.} We have
\begin{eqnarray}
\text{tr}([\Pi_i,\Pi_j B]^\dagger[\Pi_a, \Pi_b B])&=&\text{tr}\Big((\Pi_i \Pi_j B-\Pi_j B \Pi_i)^\dagger(\Pi_a \Pi_b B-\Pi_b B \Pi_a)\Big)\nonumber \\
&=&\text{tr}\Big((B\Pi_j \Pi_i -\Pi_i B \Pi_j)(\Pi_a \Pi_b B-\Pi_b B \Pi_a)\Big)\nonumber \\
&=&\text{tr}\Big(B^2 \Pi_j \Pi_i \Pi_a \Pi_b-\Pi_i B \Pi_j \Pi_a \Pi_b B-B \Pi_j \Pi_i \Pi_b B \Pi_a+\Pi_i B \Pi_j \Pi_b B \Pi_a\Big)\nonumber \\
&=&\text{tr}\Big(\underbrace{\delta_{ja} \delta_{ia}\delta_{ab} B^2 \Pi_i}_A -\underbrace{\delta_{ja} \delta_{ab} B \Pi_a B \Pi_i}_B-\underbrace{\delta_{ji} \delta_{jb} B \Pi_j B \Pi_a}_C  +\underbrace{\delta_{ia}\delta_{jb} B \Pi_j B \Pi_a}_D\Big)
\end{eqnarray}
Now we note that terms $A+B$ can be written as
\begin{eqnarray}
\text{tr}\Big(\delta_{ja} \delta_{ia}\delta_{ab} B^2 \Pi_i - \delta_{ja}\delta_{ab}B \Pi_a B \Pi_i\Big)&=&\text{tr}\Big(\delta_{ja} \delta_{ia}\delta_{ab} B^2 \Pi_i - \delta_{ja}\delta_{ab}B \Pi_a B \Pi_i\Big)
\\
&=&\delta_{ja}\delta_{ab}\text{tr}\Big( B B\Pi_i \Pi_a - B \Pi_a B \Pi_i\Big)\nonumber \\
&=&\delta_{ja}\delta_{ab}\text{tr}\Big( B (B\Pi_i \Pi_a - \Pi_a B \Pi_i)\Big)\nonumber \\
&=&\delta_{ja}\delta_{ab}\text{tr}\Big( B [B\Pi_i, \Pi_a]\Big)\nonumber \\
\end{eqnarray}
Let us now consider the terms $C+D$. We have that these can be written as
\begin{eqnarray}
(\delta_{ia}\delta_{jb} -\delta_{ji} \delta_{jb})\text{tr}( B \Pi_j B \Pi_a)=(\delta_{ia}\delta_{jb} -\delta_{ji} \delta_{jb})|b_{ja}|^2
\end{eqnarray}
which is the final result. $\square$

\textbf{Corollary 1} We have
\begin{eqnarray}
 \|[\Pi_i,\Pi_j B]\|_F^2=\delta_{ij} \text{tr}\Big(B [B\Pi_i, \Pi_i]\Big)+(1-\delta_{ij}) \text{tr}(B \Pi_jB \Pi_i)
\end{eqnarray}
\textit{Proof.} From Proposition 4 we set $a=i$ and $b=j$.
%{below has been double checked}

\textbf{Corollary 2} We have
\begin{eqnarray}
\sum_{ij} \|[\Pi_i,\Pi_j B]\|_F^2=2 \mathbb C_{\Pi}(B)
\end{eqnarray}
\textit{Proof.} From Corollary 1, we 
 have
\begin{eqnarray}
\sum_{ij} \|[\Pi_i,\Pi_j B]\|_F^2&=&\sum_{ij} \Big(\delta_{ij}\text{tr}\Big( B [B\Pi_i, \Pi_i]\Big)+(1 -\delta_{ji} )\text{tr}( B \Pi_j B \Pi_i)\Big)  \nonumber \\
&=&\sum_{ij} \delta_{ij}\text{tr}\Big( B [B\Pi_i, \Pi_i]\Big)+\sum_{ij}(1 -\delta_{ji} )\text{tr}( B \Pi_j B \Pi_i)  \nonumber \\
&=&\sum_{i} \text{tr}\Big( B [B\Pi_i, \Pi_i]\Big)+\sum_{ij} |b_{ij}|^2-\sum_i |b_{ii}|^2\nonumber \\
&=&\sum_{i} \text{tr}\Big( B [B\Pi_i, \Pi_i]\Big)+\mathbb C_{\Pi}(B) \nonumber \\
&=&2\mathbb C_{\Pi}(B)
\end{eqnarray}
where we used Eq.~(\ref{eq:cohtr}). The final equation is what we claim in the statement above. $\square$

%{below it is double checked}

{In the main text, we have provided a proof of}

\textbf{Lemma 1 - Single Normal Coherence Inequality}
\begin{eqnarray}
\|[U,A]\|_F^2  &\leq& 4 \| A\|^2\sum_{i\neq j} |a_{ij}|^2=4  \| A\|^2\mathbb{C}_U (A).
\end{eqnarray}

{Lemma 1 implies the following Corollary,}

\textbf{Corollary} For a unitary operator $U$, we have
\begin{eqnarray}
\|[U,A]\|_F^2  &\leq& 4 \sum_{i\neq j} |a_{ij}|^2=4  \mathbb{C}_U (A)
\end{eqnarray}
\textit{Proof.} It follows from Lemma 1 and the fact that $\| U\|=1$.

\textbf{Lemma $1^\prime$ - Single Unitary coherence inequality}

Let $U$ be a unitary operator and $A$ an Hermitian operator.
Then
\begin{eqnarray}
\|[U,A]\|_F^2  &\leq& 2\sqrt{2}  \|A\|\Big(2^{-\frac{1}{4}}\sqrt{n \|A\|} \mathbb C^2_U(A)+  \mathbb C^1_U(A)\Big)
\end{eqnarray}

\textit{Proof.}
Let $U=\sum_{i} \eta_i \Pi_i$, with $\Pi_{i}\Pi_j=\delta_{ij} \Pi_i$. Then,
we have 
\begin{eqnarray}
\|[U,A]\|_F^2&=& \sum_{ij} \eta_i \eta_j^* \text{tr}([\Pi_j,A]^\dagger [\Pi_i,A]) \nonumber \\
&=&-\sum_{ij} \eta_i \eta_j^* \text{tr}\Big((\Pi_j A-A \Pi_j)(\Pi_i A-A \Pi_i)\Big)\nonumber \\
&=&-2\sum_{ij} \eta_i \eta_j^* \text{tr}(\Pi_j A \Pi_i A- \delta_{ij} A^2 \Pi_i) \nonumber \\
&=&2\sum_{ij} \eta_i \eta_j^* \text{tr}(A[A\Pi_j  ,  \Pi_i]) \nonumber \\
\end{eqnarray}
We can now upper bound the  quantity above with the absolute value, and using the identity $|\text{tr}(A[A\Pi_j  ,  \Pi_i])|=|\text{tr}(A[\Pi_j A ,  \Pi_i])|$, we have
\begin{eqnarray}
\|[U,A]\|_F^2 &\leq &2\sum_{ij} |\text{tr}(A[\Pi_j A ,  \Pi_i])| \nonumber \\
&=& 2\sum_{ij} |\text{tr}(A[\Pi_j A ,  \Pi_i])| \nonumber \\
&\leq&2 \|A\| \sum_{ij} \sqrt{ \text{r}([\Pi_j A ,  \Pi_i]) \|[\Pi_j A ,  \Pi_i]\|_F^2 }
\end{eqnarray}
We note now that $\text{r}([\Pi_j A ,  \Pi_i])\leq 2$. Then
\begin{eqnarray}
\|[U,A]\|_F^2  &\leq&2\sqrt{2}  \|A\| \sum_{ij} \sqrt{  \|[\Pi_j A ,  \Pi_i]\|_F^2 }
\end{eqnarray}
We now use the fact that, from Corollary 1, 
\begin{eqnarray}
\|[U,A]\|_F^2  &\leq&2\sqrt{2}  \|A\| \sum_{ij} \sqrt{ \delta_{ij} \text{tr}\Big(A [A\Pi_i, \Pi_i]\Big)+(1-\delta_{ij}) \text{tr}(A \Pi_jA \Pi_i)}\nonumber \\
&=&2\sqrt{2}  \|A\|\Big(\sum_{i}\sqrt{\text{tr}\Big(A [A\Pi_i, \Pi_i]\Big)}+\sum_{i\neq j} \sqrt{\text{tr}(A \Pi_jA \Pi_i)}\Big)\nonumber \\
&=&2\sqrt{2}  \|A\|\Big(\sum_{i}\sqrt{\text{tr}\Big(A [A\Pi_i, \Pi_i]\Big)}+\sum_{i\neq j} |a_{ij}|\Big)\nonumber \\
&\leq& 2\sqrt{2}  \|A\|\Big(\sqrt{\sqrt{2} \|A\|}\sum_{i}\sqrt{\frac{1}{2} \|[\Pi_i,A]\|_F^2}+ \sum_{i\neq j} |a_{ij}|\Big)\nonumber \\
&\leq& 2\sqrt{2}  \|A\|\Big(2^{-\frac{1}{4}}\sqrt{\|A\|}\sum_{i}\sqrt{\ \|[\Pi_i,A]\|_F^2}+ \sum_{i\neq j} |a_{ij}|\Big)\nonumber \\
&=& 2\sqrt{2}  \|A\|\Big(2^{-\frac{1}{4}}\sqrt{\|A\|} \sum_{i} (\sum_{k\neq i} |a_{ik}|^2)^{\frac{1}{2}}+ \sum_{i\neq j} |a_{ij}|\Big)\nonumber \\
\end{eqnarray}
The bound above is the first step of the bound.
We now use the fact that the square root is concave, and thus $\sum_{ij} \sqrt{\lambda_{ij} a_{ij}}\leq \sqrt{\sum_{ij} \lambda_{ij}}\sqrt{\sum_{ij} a_{ij}}$. Then
\begin{eqnarray}
\|[U,A]\|_F^2  &\leq& 2\sqrt{2}  \|A\|\Big(2^{-\frac{1}{4}}\sqrt{\|A\|} \sqrt{n} \sqrt{ \sum_{i\neq j} |a_{ij}|^2}+ \sum_{i\neq j} |a_{ij}|\Big)
\end{eqnarray}
It follows that
\begin{eqnarray}
\|[U,A]\|_F^2  &\leq& 2\sqrt{2}  \|A\|\Big(2^{-\frac{1}{4}}\sqrt{n \|A\|} \mathbb C^2_U(A)+  \mathbb C^1_U(A)\Big)
\end{eqnarray}
which concludes the proof.  $\square$

\textbf{Lemma 2 - Double Normal Coherence Inequality}  Assume A be a normal operator and let B be Hermitian. Then, 
$$\|[A^\dagger,BA]\|_F^2\leq 4\| A\|^4 \mathbb{C}_A(B)$$

\textit{Proof}. 
For simplicity, we assume that $A$ is a normal operator, e.g.  we have $A=\sum_i \Pi_i \eta_i$, where $\eta_i$ are the eigenvalues of $A$. If the operator is normal, then $A$ and $A^\dagger$ have an identical spectral decomposition and the proof goes along similar steps, which is Lemma 3.

Then
\begin{eqnarray}
\|[A^\dagger,BA]\|_F^2&=&\|\sum_{ij} \eta_i^* \eta_j [\Pi_i,B\Pi_j]\|_F^2=\sum_{ij}\sum_{ab}\eta_i \eta_j^* \eta_a^* \eta_b\text{tr}([\Pi_i,B \Pi_j]^\dagger [\Pi_a,B \Pi_b]).\nonumber \\
\end{eqnarray}
Now using Proposition 4, we have
\begin{eqnarray}
\|[A,BA]\|_F^2&=&\|\sum_{ij} \eta_i \eta_j [\Pi_i,B\Pi_j]\|_F^2=\sum_{ij} \sum_{ab}\eta_i \eta_j^* \eta_a^* \eta_b \text{tr}([\Pi_i,B\Pi_j]^\dagger [\Pi_a,B\Pi_b])\nonumber \\
&=&\sum_{ij}\sum_{ab}\eta_i \eta_j^* \eta_a^* \eta_b\Big(\delta_{ja}\delta_{ab}\text{tr}\Big( B [B\Pi_i, \Pi_a]\Big)+(\delta_{ia}\delta_{jb} -\delta_{ji} \delta_{ib})\text{tr}( B \Pi_j B \Pi_a)\Big)\nonumber \nonumber \\
&=&\sum_{ij}\eta_i \eta_j^* |\eta_j|^2\text{tr}\Big( B [B\Pi_i, \Pi_j]\Big)+\sum_{i \neq j} (|\eta_i|^2 |\eta_j|^2-|\eta_i|^2 \eta_i \eta_j^*) | |b_{ij}|^2 \nonumber \\
&=&\sum_{ij}\eta_i \eta_j^* |\eta_j|^2 \delta_{ij} \text{tr}\Big( B^2 \Pi_i- B\Pi_i B \Pi_i\Big)-\sum_{i\neq j}\eta_i \eta_j^* |\eta_j|^2 \text{tr}\Big( B \Pi_i B \Pi_j\Big)+\sum_{i  \neq j} \eta_i\eta_j^*(\eta_i^* \eta_j-|\eta_i|^2) | |b_{ij}|^2  \nonumber \\
&\leq &\text{sup}_k |\eta_k|^4 \sum_{i} \text{tr}\Big( B^2 \Pi_i- B\Pi_i B \Pi_i\Big)+\sum_{i  \neq j} |\eta_i\eta_j^*(\eta_i^* \eta_j-|\eta_i|^2-|\eta_j|^2)| | |b_{ij}|^2  \nonumber \\
&\leq &\text{sup}_k |\eta_k|^4 \sum_{i\neq j} | |b_{ij}|^2 +\sum_{i  \neq j} |\eta_i\eta_j(\eta_i^* \eta_j-|\eta_i|^2-|\eta_j|^2)| | |b_{ij}|^2  \nonumber \\
&\leq & 4\ \text{sup}_k |\eta_k|^4 \sum_{i\neq j} | |b_{ij}|^2 = 4\| A\|^4 \mathbb{C}_A(B)
\end{eqnarray}
which is what we stated we would prove. $\square$

There is another inequality that can be obtained for a normal operator, in terms of mixed $l_1$ coherence. We report it below for a Hermitian operator, but a similar bound applies to general normal operator.

\textbf{Lemma 2(b)  - Double Hermitian Coherence Inequality}  Assume $A$ and $B$ are Hermitian operators. Then, 
$$\|[A,BA]\|_F^2\leq \|A^2\|^2\Big(\|B\|\big( 2^{-\frac{1}{4}}\sqrt{\|B\|} \sqrt{n} \sqrt{\sum_{k\neq i} |b_{ik}|^2}+ \sum_{i\neq j} |b_{ij}|\big)+2 \mathbb C_A(B)\Big).$$
where $b_{ij}$ are the elements of the matrix $B$ in the basis of $A$.\\
\textit{Proof}. 
For simplicity, we assume that $A$ is Hermitian, we have $A=\sum_i \Pi_i \eta_i$, where $\eta_i$ are the eigenvalues of $A$ and are real. If the operator is normal, then $A$ and $A^\dagger$ have an identical spectral decomposition and the proof goes along similar steps, which is Lemma 3.

Then
\begin{eqnarray}
\|[A,BA]\|_F^2&=&\|\sum_{ij} \eta_i \eta_j [\Pi_i,B\Pi_j]\|_F^2=\sum_{ij}\sum_{ab}\eta_i \eta_j \eta_a\eta_b\text{tr}([\Pi_i,B \Pi_j]^\dagger [\Pi_a,B \Pi_b]).\nonumber \\
\end{eqnarray}
Now using Proposition 4, we have
\begin{eqnarray}
\|[A,BA]\|_F^2&=&\|\sum_{ij} \eta_i \eta_j [\Pi_i,B\Pi_j]\|_F^2=\sum_{ij} \sum_{ab} \eta_i \eta_j \eta_a \eta_b \text{tr}([\Pi_i,B\Pi_j]^\dagger [\Pi_a,B\Pi_b])\nonumber \\
&=&\sum_{ij}\sum_{ab}\eta_i \eta_j \eta_a\eta_b\Big(\delta_{ja}\delta_{ab}\text{tr}\Big( B [B\Pi_i, \Pi_a]\Big)+(\delta_{ia}\delta_{jb} -\delta_{ji} \delta_{ib})\text{tr}( B \Pi_j B \Pi_a)\Big)\nonumber \nonumber \\
&=&\sum_{ij}\eta_i \eta_j |\eta_j|^2\text{tr}\Big( B [B\Pi_i, \Pi_j]\Big)+\sum_{i  j} (\eta_i^2 \eta_j^2-\eta_i^3\eta_j) | |b_{ij}|^2 \nonumber \\
&=&\sum_{ij}\eta_i \eta_j |\eta_j|^2\text{tr}\Big( B [B\Pi_i, \Pi_j]\Big)+\sum_{i  \neq j} \eta_i\eta_j(\eta_i \eta_j-\eta_i^2) | |b_{ij}|^2  \nonumber \\
&\leq& \sum_{ij}|\eta_i \eta_j|\ |\eta_j|^2\ |\text{tr}\Big( B [B\Pi_i, \Pi_j]\Big)|+\sum_{i  \neq j} |\eta_i\eta_j(\eta_i \eta_j-\eta_i^2)|\ | |b_{ij}|^2
\end{eqnarray}
Here we wrote $b_{ij}$ as the elements $ij$ of the operator $B$ in the basis of $\Pi$'s.
Let us now look at the two expressions separately. The exact expression for first term is given by Frobenius norm is thus bounded by another Frobenius norm as follows:
First, we note that $|\text{tr}([B[B\Pi_i,\Pi_j])|=|\text{tr}([B[\Pi_iB,\Pi_j])|$. Then, using $|\text{tr}(A B)|\leq \|A\|\sqrt{ \text{r}(B)} \|B\|_F$,
\begin{eqnarray}
\sum_{ij}|\eta_i \eta_j|\ |\eta_j|^2\ |\text{tr}\Big( B [B\Pi_i, \Pi_j]\Big)|&=&\sum_{ij} |\eta_i\eta_j| \eta_i^2 \text{tr}([B[\Pi_iB,\Pi_j])|\leq\|B\|\sum_{ij} |\eta_i\eta_j| \eta_i^2 \sqrt{  \text{r}([\Pi_iB,\Pi_j]) \|[\Pi_iB,\Pi_j]]\|_F^2} \nonumber \\
\end{eqnarray}
Now note that $\text{r}([\Pi_i B,\Pi_j)=\text{r}(\Pi_i B\Pi_j-\Pi_j \Pi_i B)\leq 2$, and using Propositions 1, 2 and 3, we have 
\begin{eqnarray}
\|[\Pi_iB,\Pi_j]]\|_F^2=\delta_{ij} \text{tr}\Big(B [B\Pi_i, \Pi_i]\Big)+(1-\delta_{ij}) \text{tr}(B \Pi_jB \Pi_i).
\end{eqnarray}
The first term can then be bounded by
\begin{eqnarray}
\sum_{ij}|\eta_i \eta_j|\ |\eta_j|^2\ |\text{tr}\Big( B [B\Pi_i, \Pi_j]\Big)|&\leq&\|B\| \sum_{ij} |\eta_i|^4 \sqrt{\delta_{ij} \text{tr}\Big(B [B\Pi_i, \Pi_i]\Big)+(1-\delta_{ij}) \text{tr}(B \Pi_jB \Pi_i)}\nonumber \\
&=&\|B\|\ \|A^2\|^2  \Big(\sum_{i}\sqrt{\text{tr}\Big(B [B\Pi_i, \Pi_i]\Big)}+\sum_{i\neq j} \sqrt{\text{tr}(B \Pi_jB \Pi_i)}\Big) \nonumber \\
&=&\|B\|\ \|A^2\|^2  \Big(2^{-\frac{1}{4}}\sqrt{\|B\|} \sum_{i} (\sum_{k\neq i} |b_{ik}|^2)^{\frac{1}{4}}+ \sum_{i\neq j} |b_{ij}|\Big)\nonumber \\
&\leq &\|B\|\ \|A^2\|^2  \Big(2^{-\frac{1}{4}}\sqrt{\|B\|} \sqrt{n} \sqrt{\sum_{k\neq i} |b_{ik}|^2}+ \sum_{i\neq j} |b_{ij}|\Big)
 \nonumber \\
\end{eqnarray}
For the second term, we have
\begin{eqnarray}
\sum_{i  \neq j}| \eta_i\eta_j(\eta_i \eta_j-\eta_i^2) |\ |b_{ij}|^2 \leq \text{sup}_{ij} |\eta_i\eta_j(\eta_i \eta_j-\eta_i^2)| \sum_{i\neq j} |b_{ij}|^2.
\end{eqnarray}
Now note that
\begin{eqnarray}
\text{sup}_{ij} |\eta_i\eta_j(\eta_i \eta_j-\eta_i^2)|\leq 2 \|A^2\|^2
\end{eqnarray}
while $\sum_{i\neq j} |b_{ij}|^2=\frac{1}{2}\mathbb C_{A}(B)$.
The second term can then be bounded by 
\begin{eqnarray}
\sum_{i  \neq j} |\eta_i\eta_j(\eta_i \eta_j-\eta_i^2)| |b_{ij}|^2\leq 2 \|A^2\|^2 \mathbb C_A(B)
\end{eqnarray}
We thus obtain the final bound
\begin{eqnarray}
\|[A,BA]\|_F^2\leq \|A^2\|^2\Big(\|B\|\big( 2^{-\frac{1}{4}}\sqrt{\|B\|} \sqrt{n} \sqrt{\sum_{k\neq i} |b_{ik}|^2}+ \sum_{i\neq j} |b_{ij}|\big)+2 \mathbb C_A(B)\Big)
\end{eqnarray}
which is what we stated we would prove. $\square$

\end{document}